\documentclass[aps,prb,twocolumn,superscriptaddress]{revtex4}
\usepackage{amssymb}
\usepackage{graphicx}
\usepackage{amsmath}

\usepackage{epsf}

\begin{document}

\title{Replica theory for fluctuations of the activation barriers in glassy
systems}
\author{Maxim Dzero}
\affiliation{Center for Materials Theory, Rutgers University, Piscataway, NJ 08854, USA}
\author{ J\"{o}rg Schmalian}
\affiliation{Department of Physics and Astronomy and Ames Laboratory, Iowa State
University, Ames, IA 50011}
\author{Peter G. Wolynes}
\affiliation{Department of Chemistry and Biochemistry and Department of Physics,
University of California, San Diego, La Jolla, CA 92093 }
\date{\today}

\begin{abstract}
We consider the problem of slow activation dynamics in glassy systems undergoing 
a random first order phase transition. Using an effective potential approach to supercooled liquids, 
we determine the spectrum of activation barriers for entropic droplets. 
We demonstrate that fluctuations of
the configurational entropy \ and of the liquid glass surface tension are
crucial to achieve an understanding of the barrier fluctuations in glassy
systems \ and thus are ultimatively responsible for the broad spectrum of
excitations and heterogeneous dynamics in glasses.  In particular we derive
a relation between the length scale for dynamic heterogeneity and the
related barrier fluctuations. Diluted entropic droplets are shown to have a
Gaussian distribution of barriers, strongly suggesting that non-Gaussian
behavior results from droplet-droplet interactions. 
\end{abstract}

\maketitle

\section{Introduction}

The glass transition and glassy behavior are dynamical phenomena
characterized by slow relaxations and a broad spectrum of excitations.
Recently strong experimental evidence for dynamical heterogeneity in glassy
systems, i.e. for spatially varying characteristic time scales for
relaxation, has been obtained through NMR \cite{Spiess} and
nanometer-scale probing of dielectric fluctuations\cite{Russel00}. The wide
distribution of excitations, as seen in bulk dielectric measuments, is
therefore believed to be caused largely by spatial variation of the characteristic
time scales; see also Ref.\cite{Richert02}. Regions in \ a glassy system
separated by only a few nanometers relax on time scales different by many
orders of magnitude.

\ Whether these purely dynamical phenomena can be explained in terms of the
underlying \emph{energy landscape} is still debated. The question has been
positively answered on the level of the mean field theory of glasses. The
dynamical, ideal mode coupling theory\cite{mc} and energy landscape based
replica mean field theories were demonstrated to describe the same
underlying physics, yet from rather different perspectives\cite{KT87,KW87}.
The idealized version of the mode coupling theory\cite{mc} falls out of
equilibrium below a temperature $T_{A}>T_{g}$, where $T_{g}$ is the
laboratory glass temperature (see below). This is consistent with static,
energy landscape based mean field theories that find an emergence of
exponentially many metastable states, $\mathcal{N}_{ms}\propto \exp \left(
const.V\right) $, at the same temperature $T_{A}$. Here $V$ is the volume of
the system. Thus, the configurational entropy $S_{c}=k_{B}\log \mathcal{N}%
_{ms}$ becomes extensive and only vanishes at the Kauzmann tempeature $%
T_{K}<T_{A}$. Unfortunately, neither the emergence of slow activated
dynamics, nor the possibility of spatial heterogeneity are correctly
captured within mean field theory, making it impossible to use mean field theory
alone to directly explain
the heterogeneous dynamics observed in numerous glass forming liquids. On
the other hand, mode coupling theory at higher temperatures ($T>T_{A}$) has
been shown to give a reasonably accurate account of the dynamics of liquids.
At lower temperatures ($T<T_{A}$), bulk thermodynamic properties of
glassforming liquids, most notably the temperature dependence of the
configurational entropy and the related Kauzmann paradox\cite{Angell98}, \
agree with the results of the energy landscape based mean field theory (see
Refs.\cite{KTW89,MP00}). This motivated the 
development of a theory for glassforming liquids that has its foundation in the
mean field theory, but that does take into account activated dynamics and
spatial heterrogeneity which are beyond the strict mean field approach\cite%
{KW87,KTW89}. The resulting random first order \ (RFOT) theory of glasses demonstrates
that ergodicity above the Kauzmann temperature $T_{K}$ can be restored via
\textquotedblleft entropic droplets\textquotedblright , in which a region of
the liquid is replaced by any of an exponentially large number of
alternatives. The entropy of the possible alternatives acts as a driving
force for structure change. This driving force is off-set by the free energy
cost of matching two alternative structures at their boundaries. This
conflict gives a free energy barrier for activated motions\cite%
{KW87,Biroli04}. Various consequences of entropic droplets have been analyzed in
Refs.\cite{XW00,XW01,Lubchenko01,Lubchenko04a} using for concrete calculation
a density functional description of the liquid state. These quantitative results 
have been found to be in good agreement
with experiment. Within RFOT, the number of metastable states determines the
activation barrier of the glassy state. While this is similar to the
Adam-Gibbs theory for glasses\cite{Adam65}, there exist important
distinctions between the two approaches, see Ref.\cite{Biroli04}. 
In agreement with the RFOT approach but not with the assumptions of 
Adams-Gibbs argument the complexity of a correlated region increases as the 
glass transition is approached \cite{Capaccioli08}.
The need for instanton like events to understand
the long time dynamics in glasses is also supported by recent findings
extended mode-coupling theories for dense fluids\cite{Cates06}.

More recently, an effective Landau theory for glasses, based on the replica
method of Ref.\cite{Franz95}, has been developed. This framework naturally
allows for activated dynamics beyond mean field theory\cite%
{Franz05-1,Dzero05}. This approach based on instantons offers a formal
justification of the entropic droplet approach using the replica approach to
glasses. It reproduces and thus confirms several results of the random first
order transition (RFOT) theory of structural glasses\cite{KTW89}. In both
approaches, a mean droplet activation energy $\overline{F^{\ddagger }}$
occurs that is determined by the configurational entropy density $s_{c}$
(for details see below): 
\begin{equation}
\overline{F^{\ddagger }}\propto s_{c}^{1-d\nu }.
\end{equation}%
$s_{c}\propto \frac{T-T_{K}}{T_{K}}$ vanishes linearly at the Kauzmann
temperature $T_{K}<T_{g}$. The exponent $\nu $ relates the droplet size $R$
and the configurational entropy density: $R\propto s_{c}^{-\nu }$. $d$ is
the spatial dimension. Below we assume that $d=3$. In its most elementary
version the replica Landau theory yields $\nu =1$\cite{Franz05-1,Dzero05}. A
renormalization of the droplet interface due to wetting of intermediate
states on the droplet surface \cite{KTW89} was shown to yield $\nu =2/d$,
leading to $\overline{F^{\ddagger }}\propto Ts_{c}^{-1}$ and correspondingly
to a Vogel-Fulcher law $\overline{\tau }\simeq \tau _{0}\exp \left( \frac{%
DT_{K}}{T-T_{K}}\right) $ for the mean relaxation time 
\begin{equation}
\overline{\tau }=\tau _{0}\exp \left( \frac{\overline{F^{\ddagger }}}{k_{B}T}%
\right) .  \label{meanr}
\end{equation}%
A softening of the surface tension due to replica symmetry breaking of the
instanton solution was also obtained in Ref. \cite{Dzero05} and seems to be
a first correction containing the effects that lead to a reduction of the exponent $\nu $
from the value $\nu =1$.

Numerous experiments on supercooled liquids not only depend on the
mean barrier, $\overline{F^{\ddagger }}$, but are sensitive to the
entire, broad excitation spectrum in glasses\cite{Richert02}. Most notably,
the broad peaks in the imaginary part of the dielectric function $%
\varepsilon ^{\prime \prime }\left( \omega \right) $ are most natually
understood in terms of a distribution $g\left( \tau \right) $ of relaxation
times, \ such that 
\begin{equation}
\varepsilon ^{\prime \prime }\left( \omega \right) \propto \int d\tau
g\left( \tau \right) \frac{\omega \tau }{1+\left( \omega \tau \right) ^{2}}.
\end{equation}%
Similarly dynamical heterogeneity with spatially fluctuating relaxation
times yields nonexponential (frequently streched exponential) relaxation of
the correlation function 
\begin{equation}
\phi \left( t\right) =\int d\tau g\left( \tau \right) e^{-t/\tau },
\label{phi}
\end{equation}%
and allows for an explanation of nonresonant hole burning experiments\cite%
{Diezemann01}, even though interesting interpretations of the non-exponentiality 
based on "dynamical homogeneity" exist as well\cite{Cugliandolo00}. Other effects that are most
likely caused by a distribution of relaxation rates include the breakdown
of the Stokes-Einstein relation $D=\frac{k_{B}T}{4\pi \eta L}$ between the
diffusion coefficient $D$ of a particle of size $L$ and the viscosity $\eta $%
\cite{Fujara92,Cicerone95}.

These experiments all call for a more detailed analysis of the fluctuations 
\begin{equation*}
\overline{\delta F^{\ddagger 2}}\equiv \overline{F^{\ddagger 2}}-\overline{%
F^{\ddagger }}^{2}
\end{equation*}%
of the activation barriers and, more generally, of the distribution function 
$p\left( F^{\ddagger }\right) $ of barriers. The latter yields the
distribution function of the relaxation times 
\begin{equation}
g\left( \tau \right) =p\left( F^{\ddagger }\right) \frac{dF^{\ddagger }}{%
d\tau }
\end{equation}%
through $\tau \left( F^{\ddagger }\right) =\tau _{0}\exp \left( \frac{%
F^{\ddagger }}{k_{B}T}\right) $. For example, in case of a Gaussian
distribution of barriers one obtains a broad, log-normal distribution of
relaxation rates:%
\begin{equation}
g\left( \tau \right) =\frac{1}{\tau \sqrt{2\pi \lambda }}\exp \left( -\frac{%
\log ^{2}\left( \tau /\overline{\tau }\right) }{2\lambda }\right) ,
\label{lognorm}
\end{equation}%
with 
\begin{equation}
\lambda =\overline{\delta F^{\ddagger 2}}/\left( k_{B}T\right) ^{2}
\end{equation}%
and $\overline{\tau }$ from Eq.\ref{meanr}. While the distribution, Eq.\ref%
{lognorm}, does not yield precisely a stretched exponential form for the correlation
function, it can often be approximated by 
\begin{equation}
\phi \left( t\right) \simeq \exp \left( -(t/\overline{\tau })^{\beta }\right)
\end{equation}%
with $\beta =\left( 1+\lambda \right) ^{-1/2}$. Furthermore, the study of
higher order moments of $p\left( F^{\ddagger }\right) $ is important to
determine whether the distribution is indeed Gaussian or more complicated.

In this paper we determine the fluctuations $\overline{\delta F^{\ddagger 2}}
$ as well as higher moments of the distribution of the barrier distribution
function $p\left( F^{\ddagger }\right) $. \ We start from the replica Landau
theory of Ref.\cite{Franz95} and use the replica instanton theory of Refs.%
\cite{Franz05-1,Dzero05}. We then generalize the replica formalism to
determine barrier fluctuations in addition to the determination of the mean
barrier $\overline{F^{\ddagger }}$ . Our approach is an important step to
gain insight into the whole spectrum of activated processes in glassy
systems in general.

\begin{figure}[t]
\includegraphics[width=2.8in]{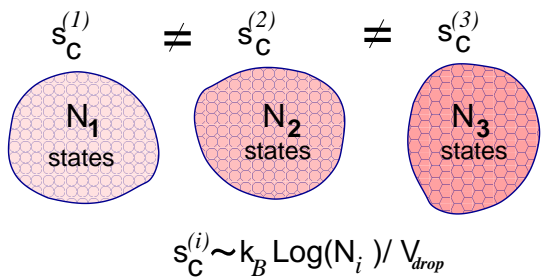}
\caption{Appearance of an exponentially large number of meta-stable
configurations and the possibility for a system to realize these
configurations gives rise to the configurational entropy. The latter serves
as the driving force for the structure change in the glassy phase. The
energy cost to match the boundaries between the different density
configurations gives rise to the energy barriers. The fluctuations of the
configurational entropy give rise to the energy barrier fluctuations.}
\label{fig1}
\end{figure}

The role of barrier fluctuations within the RFOT theory of structural
glasses was first discussed \ by Xia and Wolynes in Ref.\cite{XW01}.
Variations in the configurational entropy density were assumed to be the
main cause of the fluctuations of the barrier, Fig. \ref{fig1}. 
%%%%%%%%%% Fig. 1: droplets
%%%%%%%%%% End of Fig.1
Starting from $F^{\ddagger }\propto s_{c}^{1-d\nu }$, barrier and
configurational entropy density fluctuations are related by $\delta
F^{\ddagger }\propto s_{c}^{-d\nu }\delta s_{c}$, which yields $\overline{%
\delta F^{\ddagger 2}}\propto s_{c}^{-2d\nu }\overline{\delta s_{c}^{2}}$.
Standard fluctuation theory then determines the entropy fluctuations in a
region of size $R^{d}$ as $\overline{\delta s_{c}^{2}}\propto \Delta
c_{p}R^{-d}$ where the configurational heat capacity remains finite as $%
T\rightarrow T_{K}$ (for a detailed discussion of $\Delta c_{p}$ see below).
This reasoning finally yields%
\begin{equation}
\overline{\delta F^{\ddagger 2}}\propto s_{c}^{-d\nu }\propto R^{d}.
\label{XW}
\end{equation}%
In the approach of \ Ref.\cite{XW01}, barrier fluctuations diverge \ with
the volume of the entropic droplet. Our results will demonstrate that the
leading droplet size dependence of $\overline{\delta F^{\ddagger 2}}$ agrees
with Eq.\ref{XW} and Ref.\cite{XW01}. We also show that there are additional
terms originating in the fluctuations of the interface tension of entropic
droplets that contribute to $\overline{\delta F^{\ddagger 2}}$. In addition
we analyze the temperature dependence of the static barrier fluctuations and
of the \ exponent $\beta $ for stretched exponential relaxation as well as
the variation of the dielectric response with frequency and temperature.

This paper is organized as follows. In the next section we introduce the
replica effective potential formalism to analyze barrier fluctuations in
glassforming liquids. We motivate the approach by starting from a density
functional approach to liquids and using \ an equilibrium replica theory to
derive the mean field theory as starting point of our calculation. Next we
analyze fluctuations of the configurational entropy in bulk, summarize the
effective potential approach of Ref.\cite{Franz95} and the instanton
calculation that yields the mean barrier $\overline{F^{\ddagger }}$. Finally
we demonstrate how higher moments of the barrier distribution can be derived
within the same formalism. At the end of the section we present our results
for physical observables. The paper is concluded with a summarizing section.

\section{The replica effective potential formalism}

\subsection{Motivation and cloned replica approach}

\begin{figure}[t]
\includegraphics[width=2.8in]{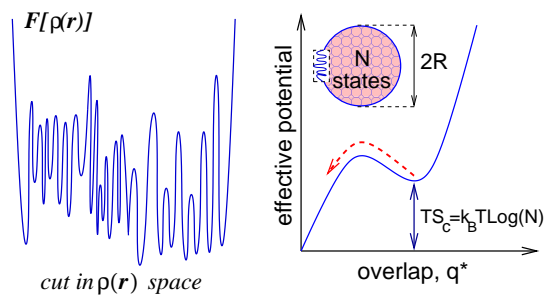}
\caption{The typical energy landscape in the structural glassy phase as a
function of density (left panel) and the effective potential, which is used
to describe the transition between the nonergodic glassy states and the
ergodic liquid state (right panel).}
\label{fig2}
\end{figure}

We start \ our description of glassy systems from the point of view that
there exists a density functional, $\phi \left[ \rho \right] $, that
describes a supercooled liquid undergoing a mean field glass transition.
Initially, such an approach was used in Ref.\cite{Singh85} where it was
shown that it allows one to describe the emergence of a metastable amorphous
solid, Fig. \ref{fig2}. Following the classical approach to freezing into
ordered crystalline states\cite{Ramakrishnan}, the density was assumed to be 
$\rho \left( \mathbf{r}\right) =\left( \frac{\alpha }{\pi }\right)
^{3/2}\sum_{i}e^{-\alpha \left( \mathbf{r-r}_{i}\right) ^{2}}$. Here,\ $%
\alpha $ determines the Lindemann length, $\alpha ^{-1/2}$ over which
particles are localized. In distinction to crystallization, the mean
positions, $\mathbf{r}_{i}$, of \ an amorphous solid were taken to be those
of a random \ hard sphere packing\cite{Bennett72}, instead of the periodic
crystal lattice positions. The free energy of this amorphous solid was shown
to have a global minimum at $\alpha =0$ and a local minimum for finite $%
\alpha $. If $\alpha \simeq V^{-2/3}\rightarrow 0$, the particles are
delocalized and the system in an ergodic liquid state with homogeneous
density $\rho _{0}=\frac{N}{V}$. Finite $\alpha $ corresponds to a frozen
amorphous solid i.e. a glassy state. For $T>T_{K}$ the amorphous solid is
metastable with respect to the liquid, and higher in free energy by $TS_{c}$%
. It is of course always metastable with respect to the crystalline solid. 
%%%%%%%%%% Fig2: Schematic plot of barriers and relation to eff. potential
%%%%%%%%%% End of Fig2
Essentially the same mean field physics can be described \ in terms of a
formally more precise replica approach, introduced by Monasson\cite{Mon95}.
Here one determines the partition function in the presence of a bias
configuration $\widehat{\rho }\left( r\right) $:%
\begin{equation}
Z\left[ \widehat{\rho }\right] =\int D\rho e^{-\beta \phi \left[ \rho \right]
-g\int d^{d}x\left( \rho \left( x\right) -\widehat{\rho }\left( x\right)
\right) ^{2}}
\end{equation}%
Here, $\int D\rho ...$ corresponds to the statistical sum over all density
configurations of the system. The free energy of a bias configuration is 
\begin{equation}
f\left[ \widehat{\rho }\right] =-T\log Z\left[ \widehat{\rho }\right] .
\end{equation}%
Physically $f\left[ \widehat{\rho }\right] $ can be interpreted as the free
energy for a metastable amorphous configuration of atoms, for example with
density $\widehat{\rho }\left( x\right) $ being a sum over Gaussians
discussed above. In the replica formalism, no specific configuration like
this needs to be specified in order to perform the calculation. Rather, the
assumption is made that the probability distribution for metastable
configurations is determined by $f\left[ \widehat{\rho }\right] $ according 
\begin{equation}
P\left[ \widehat{\rho }\right] \propto \exp \left( -\beta _{\mathrm{eff}}f%
\left[ \widehat{\rho }\right] \right)  \label{distrms}
\end{equation}%
and is characterized by the effective temperature $T_{\mathrm{eff}}=\beta _{%
\mathrm{eff}}^{-1}\geq T$. \ This allows one to determine the mean free energy 
\begin{equation}
\overline{F}=\int D\widehat{\rho }f\left[ \widehat{\rho }\right] P\left[ 
\widehat{\rho }\right]
\end{equation}%
and the corresponding mean configurational entropy 
\begin{equation}
\overline{S_{c}}=-\int D\widehat{\rho }P\left[ \widehat{\rho }\right] \log P%
\left[ \widehat{\rho }\right] .  \label{Sc}
\end{equation}

The approach of Ref.\cite{Mon95} was successfully used to develop a mean
field theory for\ glass formation in supercooled liquids\cite{MP991}
yielding results in detailed agreement with earlier, non-replica approaches%
\cite{Stoessel84}. \ As shown by Monasson\cite{Mon95}, the mean values $%
\overline{F}$ and $\overline{S_{c}}$ can be determined from a replicated
partition function 
\begin{equation}
Z\left[ m\right] =\int D^{m}\rho e^{-\beta \sum_{a=1}^{m}\phi \left[ \rho
_{a}\right] +g\sum_{a,b=1}^{m}\int d^{d}x\rho _{a}\left( x\right) \rho
_{b}\left( x\right) }  \label{Z(m)}
\end{equation}%
via $\overline{F}=\frac{\partial }{\partial m}mF\left( m\right) $ and $%
\overline{S_{c}}=\frac{m^{2}}{T}\frac{\partial }{\partial m}F\left( m\right) 
$ with 
\begin{equation}
F\left( m\right) =-\frac{T}{m}\log Z\left( m\right)
\end{equation}%
and replica index $m=\frac{T}{T_{\mathrm{eff}}}$.\ \ It follows from the
replicated free energy $F\left( T_{\mathrm{eff}}\right) $: 
\begin{eqnarray}
\overline{F} &=&-T_{\mathrm{eff}}^{2}\frac{\partial \left( F\left( T_{%
\mathrm{eff}}\right) /T_{\mathrm{eff}}\right) }{\partial T_{\mathrm{eff}}} 
\notag \\
\overline{S_{c}} &=&-\frac{\partial F\left( T_{\mathrm{eff}}\right) }{%
\partial T_{\mathrm{eff}}},  \label{thermo}
\end{eqnarray}%
These results are in analogy to the usual thermodynamic relations between
free energy ($F\rightarrow F\left( T_{\mathrm{eff}}\right) $), internal
energy ($U\rightarrow \overline{F}$), entropy ($S\rightarrow \overline{S_{c}}
$) and temperature ($T\rightarrow T_{\mathrm{eff}}$), see also Ref.\cite{N98}%
. If the liquid gets frozen in one of the many metastable states, the system
cannot anymore realize its configurational entropy, i.e. the mean free
energy of frozen states is $\overline{F}$, higher by $T\overline{S_{c}}$ if
compared to the equilibrium free energy of the liquid, Fig. \ref{fig2}.

If the replica solution is taken to be marginally stable, so that the lowest eigenvalue of
the fluctuation spectrum beyond mean field solution vanishes, it has been shown%
\cite{WSW03} that $T_{\mathrm{eff}}$ agrees with the result obtained from
the generalized fluctuation-dissipation theorem in the dynamic description
of mean field glasses\cite{CK93}. Typically, the assumption of marginality
is appropriate for early times right after a rapid quench from high
temperatures. \ In this case $T_{\mathrm{eff}}>T$ for $T<T_{A}$, i.e. the
distribution function of the metastable states is not in equilibrium on the
time scales where mode coupling theory or the requirement for marginal
stability applies. Above the Kauzmann temperature it is however possible also to
consider the solution where $T_{\mathrm{eff}}=T$, i.e. where the
distriburion of metastable states has equilibrated with the external heat
bath of the system. Since we are interested in a system where ergodicity
is restored for $T_{K}<T<T_{A}$ we use $T_{\mathrm{eff}}=T$. 
Technically this is a supercooled liquid and not a non-equilibrium aging glass. 
Below the Kauzmann
temperature this assumption cannot be made any longer (at least within the
mean field theory) as it leads to a negative configurational entropy,
inconsistent with the definition Eq.\ref{Sc}.

The physically intuitive analogy between effective temperature, mean energy, 
$\overline{F}$ and configurational entropy to thermodynamic relations, Eq.%
\ref{thermo}, suggest that one should analyze the corresponding \emph{%
configurational heat capacity} 
\begin{equation}
C_{c}=T_{\mathrm{eff}}\frac{\partial \overline{S_{c}}}{\partial T_{\mathrm{%
eff}}}=-m\frac{\partial \overline{S_{c}}}{\partial m}.
\end{equation}%
Using the distribution function Eq.\ref{distrms} we find, as expected, that $%
C_{c}$ is\ a measure of the fluctuations of the energy and configurational
entropy of glassy states. \ We obtain for the configurational heat capacity 
\begin{equation}
C_{c}=\frac{\overline{\left( \delta F\right) ^{2}}}{T_{\mathrm{eff}}^{2}}\ =%
\overline{\left( \delta S_{c}\right) ^{2}}.  \label{ffluctC}
\end{equation}%
where $\overline{\left( \delta F\right) ^{2}}=\overline{F^{2}}-\overline{F}%
^{2}$ and $\overline{\left( \delta S_{c}\right) ^{2}}=\overline{S_{c}^{2}}-%
\overline{S_{c}}^{2}$. Here the mean values $\overline{F}$ and $\overline{%
S_{c}}$ are determined by Eq.\ref{thermo}. The fluctuations of the
configurational entropy and frozen state energy are then determined by $%
\overline{S_{c}^{2}}=\int D\widehat{\rho }P\left[ \widehat{\rho }\right]
\log ^{2}P\left[ \widehat{\rho }\right] $ and $\overline{F^{2}}=\int D%
\widehat{\rho }P\left[ \widehat{\rho }\right] f\left[ \widehat{\rho }\right]
^{2}$, respectively. Both quantities can be expressed within the replica
formalism in terms of a second derivative of $F\left( m\right) $ with
respect to $m$. For example it follows: 
\begin{equation}
\frac{\partial ^{2}}{\partial m^{2}}mF\left( m\right) =-\frac{1}{T}\left( 
\overline{F^{2}}-\overline{F}^{2}\right) .  \label{repsecd}
\end{equation}%
It is then easy to show that Eq.\ref{ffluctC} holds. With the introduction
of the configurational heat capacity into the formalism we have a measure
for the fluctuation of the number of available metastable states from its mean value.
The analogy of these results to the usual fluctuation theory of
thermodynamic variables\cite{Landau} further suggests that $C_{c}$ also
determines fluctuations of the effective temperature with mean square
deviation: 
\begin{equation}
\overline{\left( \delta T_{\mathrm{eff}}\right) ^{2}}=T_{\mathrm{eff}%
}^{2}/C_{c}.
\end{equation}%
Since $C_{c}$ is extensive, fluctuations of intensive variables, like $T_{%
\mathrm{eff}}$, or densities, like $s_{c}=S_{c}/V$, vanish for infinite
systems. However, they become relevant if one considers finite subsystems or
small droplets. In the context of glasses this aspect was first discussed by
Donth\cite{Donth}.

Using standard techniques of many body theory \ we can proceed by expressing
the replica free energy $F\left( m\right) $ as a functional of the liquid
correlation functions of the problem (see e.g. Ref. \cite{WSW04}):%
\begin{equation}
H\left[ \chi \right] =\frac{T}{2m}\left( \text{Tr}\left( \chi _{0}^{-1}\chi
\right) +T\text{Tr}\ln \chi +\Phi \left[ \chi \right] \right) .  \label{LW}
\end{equation}%
Here the mean field result of $F\left( m\right) $ is determined by the
stationary point of the effective Hamiltonian: $\frac{\delta H}{\delta \chi }%
=0$ with replicated density-density correlation function 
\begin{equation}
\chi _{ab}\left( x,x^{\prime }\right) =\left\langle \delta \rho _{a}\left(
x\right) \delta \rho _{b}\left( x^{\prime }\right) \right\rangle
\end{equation}%
The Luttinger-Ward functional $\Phi \left[ \chi \right] $ is determined by
the nature of the density-density interaction and determines the self energy 
$\Sigma =-\frac{\delta \Phi }{\delta \chi }$, see also Refs.\cite%
{mc,Bouchaud96}. \ For example, expanding the density functional, $\phi
\left( \rho \right) =\phi \left( \rho _{0}+\delta \rho \right) $, into a
Taylor series \ in $\delta \rho $, yields to lowest order in $\delta \rho $: 
\begin{equation}
\Phi =-\frac{v_{3}^{2}}{3}\int d^{d}xd^{d}x^{\prime }\sum_{ab}\chi
_{ab}^{3}\left( \mathbf{x},\mathbf{x}^{\prime }\right)  \label{phi_KB}
\end{equation}%
where $v_{3}=\frac{1}{2}\frac{d^{3}\phi \left( \rho _{0}\right) }{d\rho
_{0}^{3}}$. In what follows we will formulate the theory in terms of the
collective variable $\chi _{ab}\left( \mathbf{x},\mathbf{x}^{\prime }\right) 
$, instead of the original density fluctuations and write 
\begin{equation}
\chi _{ab}\left( \mathbf{x},\mathbf{x}^{\prime }\right) =\chi \left( \mathbf{%
x},\mathbf{x}^{\prime }\right) \delta _{ab}+C_{ab}\left( \mathbf{x},\mathbf{x%
}^{\prime }\right)
\end{equation}%
where $C_{ab}=0$ if $a=b$. $C_{ab}\left( \mathbf{x},\mathbf{x}^{\prime
}\right) $ depends on two spatial variables $x$ and $x^{\prime }$. We
simplify the problem by assuming that its Fourier transformation, $%
C_{ab}\left( \frac{\mathbf{x}+\mathbf{x}^{\prime }}{2},\mathbf{q}\right) $,
with respect to the relative coordinates $x-x^{\prime }$ can be written as%
\begin{equation}
C_{ab}\left( \mathbf{r};\mathbf{q}\right) =q_{ab}\left( \mathbf{r}\right)
\rho _{0}S\left( \mathbf{q}\right) .
\end{equation}%
with liquid structure factor, $S\left( \mathbf{q}\right) $ and $\mathbf{r}%
=\left( \mathbf{x+x}^{\prime }\right) /2\ $ is the center of mass
coordinate. The important collective variable of our theory is $q_{ab}\left( 
\mathbf{r}\right) $, which plays the role of a spatially varying
Debye-Waller factor. Within mean field theory we expect below the
temperature $T_{A}$ that $q_{ab}=q^{\ast }\left( 1-\delta _{ab}\right) $ is
replica symmetric with $q^{\ast }$ of order unity, while $q^{\ast }$
vanishes in the equilibrium liquid state for $T>T_{A}$. Physically the
Debye-Waller factor $q^{\ast }$ contains the same information as does the 
localization parameter $\alpha ^{-1}$ discussed above. Finally, it is
important to stress that a replica symmetric approach of $q_{ab}$ in the
present formalism is equivalent to one step replica symmetry breaking in the
conventional replica language\cite{Mon95}.\ 

In order to keep our calculation transparent we will not determine $H\left[ q%
\right] $ from an explicit microscopic calculation for supercooled liquids.
\ It was demonstrated by Franz and Parisi\cite{Franz97} that this is
possible. However, in order be able to make simple calculations beyond the 
mean field theory explicitly including droplet fluctuations, we start
from a simpler Landau theory in the same universality class\cite%
{Gross85,Dzero05}: 
\begin{equation}
H=E_{0}\sum_{a,b}\int \frac{d^{3}r}{a_{0}^{3}}\left( h\left[ q_{ab}\right] -%
\frac{u}{3}\sum_{c}q_{ab}q_{bc}q_{ca}\ \right)  \label{ham1}
\end{equation}%
with 
\begin{equation}
h\left[ q_{ab}\right] =\frac{a_{0}^{2}}{2}\left( \nabla q_{ab}\right) ^{2}+%
\frac{t}{2}q_{ab}^{2}-\frac{u+w}{3}q_{ab}^{3}+\frac{y}{4}q_{ab}^{4}
\label{ham2}
\end{equation}%
and replica index $a$,$b=1,\ldots ,m$. $a_{0}$ is a length scale of the
order of the first peak in the radial distribution function of the liquid
and $E_{0}$ is a typical energy of the problem that determines the absolute
value of $T\overline{s_{c}}$. In addition the problem is determined by the
dimensionless variables $t$, $u$, $w$ and $y$, which are in principle all
temperature dependent. We assume that the primary $T$-dependence is that of
the quadratic term, where $t=\frac{T-T_{0}}{E_{0}}$. 
Making this assumption, along with the polynomial form of the Lagrangian, requires constraints 
between the other Landau parameters to ensure thermodynamic consistency from the fluctuation 
formulas. In Appendix A we give estimates for the parameters of Eqs.\ref{ham1} \ and \ref{ham2} obtained from fits to experimental data for o-terphenyl (OTP), a well studied fragile
glass forming material. Formally, Eqs.\ref{ham1} \ and \ref{ham2} can be
motivated as the Taylor expansion of \ Eq.\ref{LW} together with the
functional $\Phi $ of Eq.\ref{phi_KB}. \ In what follows we further simplify
the notation and measure all energies in units of $E_{0}$ and all length
scales in units of $a_{0}$.

The mean field analysis of this model Hamiltonian is straightforward.
Inserting a replica symmetric ansatz $q_{ab}=q^{\ast }\left( 1-\delta
_{ab}\right) $ into $H$ and minimizing w.r.t $q^{\ast }$ yields $q^{\ast }=0$
or 
\begin{equation}
q^{\ast }=\ \frac{w+\sqrt{w^{2}-4ty}}{2y}.  \label{qstar}
\end{equation}%
Nontrivial solutions exits for $t<t_{A}=\frac{w^{2}}{4y}$ with $q^{\ast
}\left( t_{A}\right) =\frac{w}{2y}$, which determines the mode coupling
temperature $T_{A}=t_{A}E_{0}+T_{0}$. Inserting $q^{\ast }$ of Eq.\ref{qstar}
into $H\left[ q\right] $ yields for the replica free energy: 
\begin{equation}
\frac{F\left( m\right) }{V\left( m-1\right) }=\frac{t}{2}q^{\ast 2}-\frac{%
w+u\left( m-1\right) }{3}q^{\ast 3}+\frac{y}{4}q^{\ast 4}.  \label{F(m)Mon}
\end{equation}%
\ The mean configurational entropy density, as determined by $\overline{s_{c}%
}=\left. \frac{1}{TV}\frac{\partial }{\partial m}F\left( m\right)
\right\vert _{m\rightarrow 1}$ is given by 
\begin{equation}
T\overline{s_{c}}=\left( \frac{t}{2}q^{\ast 2}-\frac{w}{3}q^{\ast 3}+\frac{y%
}{4}q^{\ast 4}\right) .  \label{Scmon}
\end{equation}%
Inserting $q^{\ast }$ of Eq.\ref{qstar} yields the result that $\overline{s_{c}}$
vanishes at $t_{K}=\frac{2w^{2}}{9y}$ with $q^{\ast }\left( t_{K}\right) =%
\frac{2w}{3y}$. Close to $t_{K}$ it follows that 
\begin{equation}
\overline{s_{c}}\simeq \frac{t_{K}}{4y}\left( \frac{t-t_{K}}{T_{K}}\right) \propto \frac{%
T-T_{K}}{T_{K}}
\end{equation}%
as expected. At $t_{A}$ one finds $T_{A}s_{c}\left( T_{A}\right) =\frac{w^{4}%
}{192y^{3}}$. $\ $

We next determine the configurational heat capacity of the mean field theory
discussed above. From Eq.\ref{F(m)Mon} we obtain%
\begin{equation}\label{ConfHeatCap}
C_{c}\ =\frac{2Vm}{T_{\mathrm{eff}}}\ \left( \frac{t}{2}q^{\ast 2}-\frac{%
w-u\left( 2-3m\right) }{3}q^{\ast 3}+\frac{y}{4}q^{\ast 4}\right)
\end{equation}%
If $T_{\mathrm{eff}}=T$ (corresponding \ again to the equilibrium behavior
between $T_{K}$ and $T_{A}$) it follows that $C_{c}\left( T_{K}\right) =\frac{Vu%
}{T_{K}}\frac{2}{3}\left( \frac{2w}{3y}\right) ^{3}$, so that comparing it with 
the expression for the configurational entropy (\ref{Scmon}) at $t\simeq t_K$ we can write%
\begin{equation}
\overline{S_{c}}\simeq \frac{w}{4u}C_{c}\left( t_{K}\right) \ \frac{t-t_{K}}{%
t_{K}}.
\end{equation}%
Notice that $w/4ut_K$ must equal $1/T_K$ for complete thermodynamic consistency. 
For temperatures close to but above $T_{K}$ the energy fluctuations decrease
as 
\begin{equation}
C_{c}\simeq C_{c}\left( t_{K}\right) \left( 1-\frac{8u+w}{2}\frac{t-t_{K}}{%
t_{K}}\right) \ 
\end{equation}%
As the temperature approaches $T_{A}$ the configurational heat capacity
decreases faster than linearly. It reaches a value $C_{c}\left( t_{A}\right)
=\frac{V}{T_{A}}\frac{8u-w}{96}\frac{w^{3}}{y^{3}}$ signalling that the
system becomes unstable close to $t_{A}$ for $w>8u$. \ This is analogous to
the higher order replica symmetry breaking suggested by Tarzia and Moore\cite%
{Tarzia07} on the basis of the same Landau action. 
$C_{c}\left( t\rightarrow t_{A}\right) $ approaches $C_{c}\left(
t_{A}\right) $ from above with an infinite slope:%
\begin{equation}
\frac{C_{c}\left( t\rightarrow t_{A}\right) }{C_{c}\left( t_{A}\right) }=1-%
\frac{24\left( \frac{t_{A}-t}{t_{A}}\right) ^{1/2}}{8u-w}
\end{equation}%
%
%
%
%
%%%%%%%%%% Fig3: Configurational heat capacity
\begin{figure}[tbp]
\includegraphics[width=2.8in,angle=-90]{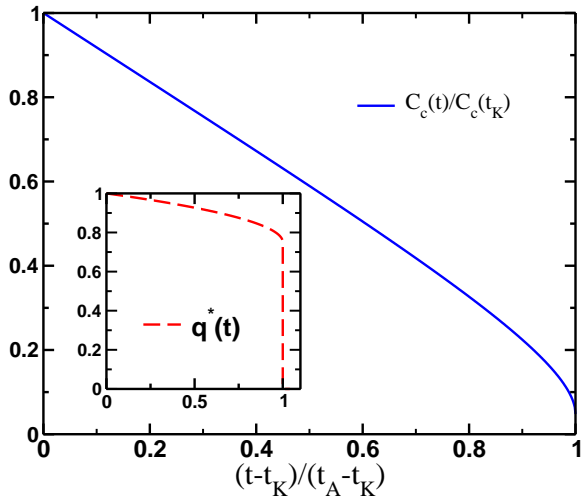}
\caption{Effective potential $\Omega \lbrack p_{c}]$ as a function of the
overlap at the fixed temperature $T_{K}<T<T_{A}$. At the Kauzmann
temperature an entropy crisis takes place with vanishing entropic advantage
of the liquid compared to the frozen solid.}
\label{fig3}
\end{figure}
%%%%%%%%%% End of Fig3

The full temperature dependences of the mean field overlap $q^*$ and
configurational heat capacity are shown in Fig. \ref{fig3}. While $q^{\ast }$
or $\overline{s_{c}}$ did not depend on the parameter $u$ of the
Hamiltonian, Eqs.\ref{ham1} \ and \ref{ham2}, the configurational heat
capacity does. We may therefore consider $u$ as a parameter that determines
the strength of entropy fluctuations.

\subsection{The effective potential approach}

In order to investigate the transition state for the decay of a specific
metastable frozen state, it is necessary to have a technique that agrees
with the mean field replica approach discussed in the previous section in
case of a homogeneous \ solution $q=q^{\ast }$, but that allows one to study the
behavior for arbitrary values of the overlap between two states, including
inhomogeneous droplet solutions. This is possible within the effective
potential approach introduced in Refs.\cite{Franz95,Franz97,Barrat97}. We
use this technique to formulate our replica instanton and barrier
fluctuation theory.

\ We are interested in the regime above $T_{K}$ (i.e. for $t>t_{K}$) and
consider the partition function of a system with constrained overlap \
between the particle density configurations $\rho _{1}\left( \mathbf{r}%
\right) $ and $\rho _{2}\left( \mathbf{r}\right) $: 
\begin{eqnarray}
Z_{p_{c},\rho _{2}} &=&\int D\rho _{1}e^{-\beta \phi \left[ \rho _{1}\right]
}  \notag \\
&&\times \prod\limits_{\mathbf{x}}\delta \left( p_{c}\left( \mathbf{r}%
\right) -\rho _{2}\left( \mathbf{r}\right) \rho _{1}\left( \mathbf{r}\right)
\right) .  \label{Zprho}
\end{eqnarray}%
The configuration $\rho _{2}$ is assumed to be in equilibrium and unaffected
by $\rho _{1}$. Averaging the free energy, $-T\log Z_{p_{c},\rho _{2}}$,
weighted with the equilibrium probability for $\rho _{2}$ yields the
corresponding mean energy for the overlap $p_{c}$: 
\begin{eqnarray}
\Omega \left[ p_{c}\right] &=&-T\left\langle \log Z_{p_{c},\rho
_{2}}\right\rangle _{\rho _{2}}  \notag \\
&=&-T\frac{\int D\rho _{2}e^{-\beta \phi \left[ \rho _{2}\right] }\log
Z_{p_{c},\rho _{2}}}{Z}.
\end{eqnarray}%
Here $Z$ is the equilibrium partition function $Z=\int D\rho e^{-\beta \phi %
\left[ \rho \right] }$. Using a replica approach $\Omega \left[ p_{c}\right] 
$ can be written as\cite{Franz95} 
\begin{equation}
\Omega \left[ p_{c}\right] =-T\lim_{n\rightarrow 0}\lim_{m\rightarrow 0}%
\frac{1}{m}\left( Z^{\left( n,m\right) }-1\right) ,
\end{equation}%
with%
\begin{eqnarray}
Z^{\left( n,m\right) } &=&\int D^{n}\rho D^{m}\widehat{\rho }e^{-\beta
\sum_{\alpha =1}^{n}\phi \left[ \rho _{\alpha }\right] -\beta
\sum_{b=1}^{m}\phi \left[ \widehat{\rho }_{b}\right] }  \notag \\
&&\times \prod\limits_{r,b=1}^{m}\delta \left( p_{c}\left( r\right) -\rho
_{1}\left( r\right) \widehat{\rho }_{b}\left( r\right) \right) .
\end{eqnarray}%
$\Omega \left[ p_{c}\right] $ can be determined by analyzing an $n+m$ times
replicated problem with additional constraint for the overlap between
certain replicas. In complete analogy to the previous paragraph one can
introduce a field theory of the overlap, but now with order parameter%
\begin{equation}
Q=\left( 
\begin{array}{cc}
r & p \\ 
p^{T} & q%
\end{array}%
\right) .
\end{equation}%
where $r$ is an $n\times n$ matrix, $q$ an $m\times m$ matrix and the $%
n\times m$ matrix $p$ must obey the additional constrained that $%
p_{1b}\left( r\right) =p_{c}\left( r\right) $ \ for $\forall _{b=1,...,m}$.
We have the same effective Hamiltonian, Eqs.\ref{ham1} and \ref{ham2} only
with $q$ replaced by $Q$, i.e.%
\begin{equation}
\overline{Z^{(n,m)}}=\int DQe^{-\beta H\left[ Q\right] }\prod%
\limits_{b=1}^{m}\delta \left( p_{1b}-p_{c}\right) .
\end{equation}%
Following Franz and Parisi\cite{Franz95} we use $r_{\alpha \beta }=0$ and
and $p_{\alpha b}=\delta _{\alpha ,1}p_{c}$ and it follows%
\begin{eqnarray}
H_{p_{c}}\left[ q\right] &=&\int d^{d}r\left( 2mh\left[ p_{c}\right]
+\sum_{ab}h\left[ q_{ab}\right] \right.  \notag \\
&&\left. -\frac{u}{3}\sum_{abc}\
q_{ab}q_{bc}q_{ca}-up_{c}^{2}\sum_{bc}q_{bc}\right) .  \label{Heffp}
\end{eqnarray}%
The remaining $q_{ab}$-dependent problem is formaly similar to the original
one of Eqs.\ref{ham1} \ and \ref{ham2}, but in an external field $up_{c}^{2}$%
.

Using these results we find that the effective potential can be written as 
\begin{equation}
\Omega \left[ p_{c}\right] =2\int d^{d}rh\left[ p_{c}\right] +\widetilde{%
\Omega }\left[ p_{c}\right]  \label{epott}
\end{equation}%
with 
\begin{equation}
\widetilde{\Omega }\left[ p_{c}\right] =-\left. \frac{\partial }{\partial m}%
\int Dqe^{-\beta H_{p_{c}}\left[ q\right] }\right\vert _{m\rightarrow 0}.
\label{intf}
\end{equation}

We start with a homogeneous replica symmetric ansatz for $q_{ab}=q\left(
\delta _{ab}-1\right) $ with equal off diagonal elements, $q$, and zero
diagonal elements. It was demonstrated that this replica symmetric ansatz is
again equivalent to one step replica symmetry breaking in the conventional
replica language\cite{comment1}. We take the limit $m\rightarrow 0$ and find 
\begin{equation}
\frac{H_{p_{c}}\left( q\right) }{Vm}=-h\left( q\right) +2h\left(
p_{c}\right) +up_{c}^{2}q-\frac{2}{3}uq^{3}.
\end{equation}%
If we extremize this with respect to $q$, i.e. perform the integration with
respect to $q$ at the level of a homogeneous saddle point approximation, it
follows that 
\begin{equation}
tq-\left( w-u\right) q^{2}+yq^{3}=up_{c}^{2}.  \label{mfclo}
\end{equation}%
Depending on the value for the overlap $p_{c}$, $q$ is in general different
from the value $q^{\ast }$ obtained within the replica approach of the
previous section. If we however insert $q$ back into $H_{p_{c}}\left(
q\right) $ and determine the effective potential $\Omega \left[ p_{c}\right]
=H_{p_{c}}\left( q\left( p_{c}\right) \right) $ we find that it has minima $%
\frac{\partial \Omega \left[ p_{c}\right] }{\partial p_{c}}=0$ for $p_{c}=0$
and for $p_{c}=q\left( p_{c}\right) =q^{\ast }$, with $q^{\ast }$ of Eq.\ref%
{qstar}, in \ complete agreement with the approach of the previous section.
The value of the effective potential for the metastable minimum $%
p_{c}=q^{\ast }$ is identical to the mean configurational entropy discussed
above 
\begin{equation}
\Omega \left[ q^{\ast }\right] =\overline{S_{c}}.
\end{equation}%
In order to analyze the local stability of the solution at the minimum of $%
\Omega \left( p_{c}\right) $ we determine the replicon eigenvalue of the
problem. Following Ref.\cite{dAT78} gives 
\begin{equation}
\lambda _{r}=t-2wq+3yq^{2}  \label{reph}
\end{equation}%
for the lowest eigenvalue. Inserting $q$ from Eq. (\ref{mfclo}) yields close
to $t_{K}$ a positive eigenvalue: $\lambda _{r}\simeq t_{K}-5\left(
t-t_{K}\right) $, while the replicon eigenvalue vanishes at $t_{A}$ as $%
\lambda _{r}\simeq 2t_{A}\left( \frac{t_{A}-t}{t_{A}}\right) ^{1/2}$. This
behavior is completely consistent with the results obtained in Ref.\cite%
{Cwilich89} for the random Potts model. The homogeneous solution is marginal
at $T_{A}$ and stable below $T_{A}$. Except for temperatures close to $T_{A}$%
, the mean field approach is locally stable with respect to small
fluctuations. %%%%%%%%%% Fig4: effective potential
\begin{figure}[tbp]
\includegraphics[width=2.8in,angle=-90]{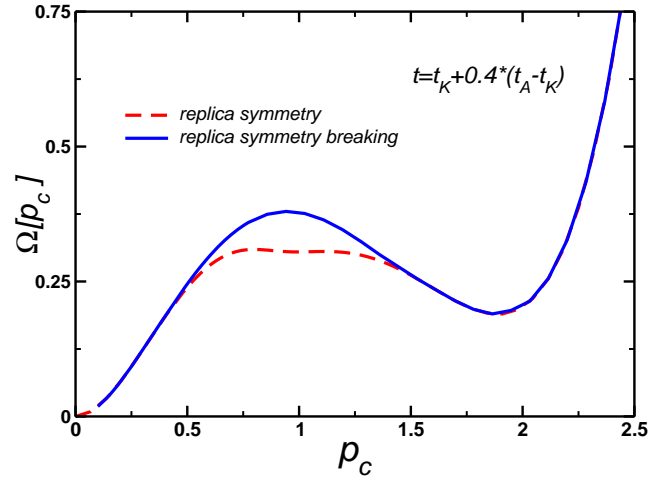}
\caption{Mean field value for the configurational heat capacity (main plot) 
and overlap (inset) for $y=1.82,w=2.72,u=0.385$ (see text) as a function of reduced
temperature $(t-t_{K})/(t_{A}-t_{K})$. Exponentially many states emerge
below $t_{A}$ with the discontinuous jump of the overlap.}
\label{fig4}
\end{figure}
%%%%%%%%%% End of Fig4
The effective potential approach is not restricted to overlaps $p_{c}$ that
minimize $\Omega \left[ p_{c}\right] $. In general $q\left( p_{c}\right) $
is given by the solution of Eq.\ref{mfclo}. \ Inserting this solution into $%
H_{p_{c}}\left( q\right) $ gives $\Omega \left( p_{c}\right) $. The result
is shown as broken line in Fig.(\ref{fig5}). It turns out that for
intermediate values of $p_{c}$, between $p_{c}=0$ and $p_{c}=q^{\ast }$ an
additional replica symmetry breaking of \ $q_{ab}$ occurs\cite{Barrat97}.
For the present model this effect was also studied in Ref.\cite{Dzero05}.
This is related to the well known effect that an external field can cause
replica symmetry breaking above $T_{K}$\cite{Sommers}, yielding one step
replica symmetry breaking for $q_{ab}$. The result for $\Omega \left(
p_{c}\right) $ shown in Fig. \ref{fig4} has been obtained, including this
additional replica symmetry breaking. Physically this additional replica
symmetry breaking as $p_{c}$ "goes over the hill" was discussed in Ref.\cite%
{Dzero05}. We also note that the generic behavior for $\Omega \left(
p_{c}\right) $ as shown in Fig. \ref{fig4} is very similar to $T\overline{%
s_{c}}\left( q^{\ast }\right) $ of Eq.\ref{Scmon}, \ if we simply plot this
function for arbitrary $q^{\ast }$. As required both expressions exactly
agree at saddle points and minima.

In summary, we showed that\ on the level of spatially homogeneous mean field
solutions, the effective potential approach of Refs.\cite%
{Franz95,Franz97,Barrat97} yields results in complete agreement with the
physically transparent approach of Ref.\cite{Mon95}. \ Since it allows for
arbitrary overlap between two states, we use it as starting point for the
determination of activated events within an instanton calculation.

\subsection{Instanton theory for activated events}

At the mean field level a glass at $T<T_{A}$ is frozen in metastable states
and characterized by the overlap $q^{\ast }$ of Eq.\ref{qstar} between
configurations at distant times. For temperatures $\ T>T_{K}$ above the
Kauzmann temperature, the free energy of such a frozen state is higher by
the configurational entropy $T\overline{S}_{c}$ compared to the liquid state
that is characterized by $q^{\ast }=0$. Thus, for $T_{K}<T<T_{A}$ the mean
field glass is locally stable ( the replicon eigenvalue $\lambda _{r}$ is
positive) but globally unstable with respect to the ergodic liquid. The
effective potential $\Omega \left( p_{c}\right) $ shown in Fig. 3 suggests
that the decay modes for the frozen state are droplet excitations, similar
to the nucleation of an unstable phase close to a first order transition.
This situation was analyzed in \ Ref.\cite{Franz05-1,Dzero05}. Using this
approach, an analysis of the effective droplet size for a Kac-type model was
performed in Ref.\cite{Franz07} In agreement with the RFOT theory\cite{KTW89}%
, the driving force for nucleation is the configurational entropy, leading
to the notion of entropic droplets.

Instanton solutions for entropic droplets are determined from%
\begin{equation}
\frac{\delta \Omega \left[ p_{c}\right] }{\delta p_{c}\left( \mathbf{r}%
\right) }=0  \label{inststat}
\end{equation}%
where we allow for spatial variations of the overlap $p_{c}\left( \mathbf{r}%
\right) $. Recently, Franz has given a transparent dynamical interpretation
for the solutions of Eq.\ref{inststat}\cite{Franz06}. Assuming that the
effective potential is characterized by spatially anisotropic but replica
symmetric solutions $q_{ab}\left( \mathbf{r}\right) =q\left( \mathbf{r}%
\right) \left( \delta _{ab}-1\right) $, we perform the integral over the $%
q_{ab}$ in Eq.\ref{intf} at the saddle point level. We find a solution $%
q\left( \mathbf{r}\right) =p_{c}\left( \mathbf{r}\right) $ with%
\begin{equation}
\nabla ^{2}p_{c}\left( \mathbf{r}\right) =\frac{dV\left( p_{c}\left( \mathbf{%
r}\right) \right) }{dp_{c}\left( \mathbf{r}\right) },  \label{inst1}
\end{equation}%
where 
\begin{equation}
V\left( p_{c}\right) =\frac{t}{2}p_{c}^{2}-\frac{w}{3}p_{c}^{3}+\frac{y}{4}%
p_{c}^{4}.
\end{equation}%
Eq.\ref{inst1} admits an exact solution in the \ thin wall limit $R\gg \xi $%: 
\begin{equation}
p_{c}(r/\xi)=q^{\ast }+\sqrt{\frac{2}{y\xi ^{2}}}\left[ \text{th}\left( \frac{r}{%
\xi }-z_{0}\right) -\text{th}\left( \frac{r}{\xi }+z_{0}\right) \right] ,
\label{pcthinwall}
\end{equation}%
where the integration constant $z_{0}$ is a function of $t,w$ and $y$. $R$
is the droplet radius and $\xi $ is the interface width given by 
\begin{equation}\label{xi}
\xi =\frac{4a_{0}}{\sqrt{3y(2q^{\ast }-q_{K}^{\ast })^{2}-6t_{K}+4t}}.
\end{equation}%
The final expression for the function $z_0(t)$ is more involved and is given in Appendix C. 

%%%%%%%%%% Fig4a: xi & R
\begin{figure}[h]
\includegraphics[width=2.8in]{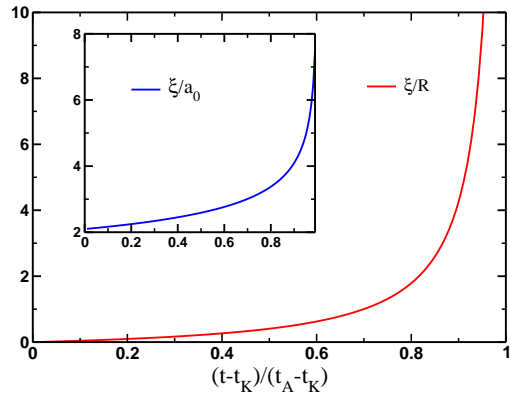}
%\centerline{\psfig{file=Fig4a_xiR.eps,height=6cm,width=8cm,angle=-90}}
\caption{We show the temperature
dependence of the ratio $\xi/R$ ($R$ is the size of the droplet) 
together with the temperature dependence of interface width $\xi(t)$ (inset).}
\label{fig5}
\end{figure}
%%%%%%%%%% End of Fig6 %%%%%%%%%%%%%%%%%%

Inserting the solution Eq.\ref{pcthinwall} into the expression for the
effective potential Eq.\ref{epott} we calculate the value of the mean
barrier. The latter is determined by optimizing the energy gain due to
creation of a droplet and energy loss due to the surface formation. As a
result for the mean barrier we find (reintroducing the energy scale $E_{0}$
and length scale $a_{0}$) 
\begin{equation}
\overline{F^{\ddagger }}=E_{0}\frac{32\pi a_{0}}{9y\xi ^{3}}R^{2},
\label{Fdagger}
\end{equation}%
The droplet radius 
\begin{equation}
R=\frac{64a_{0}^{4}}{3y^{2}q^{\ast 3}(q_{K}^{\ast }-q^{\ast }\left( t\right)
)\xi ^{3}}
\end{equation}%
is determined from the balance between the interface tension and the
entropic driving force for nucleation. The barrier energy $\overline{%
F^{\ddagger }}$ of Eq.\ref{Fdagger} determines the relaxation time $%
\overline{\tau }$ of Eq.\ref{meanr}.  Furthermore, $q_{K}^{\ast }\equiv
q^{\ast }(t=t_{K})$ is the order parameter at the Kauzmann temperature. When
temperature approaches the $T_{K}$ the radius of the droplet as well as the
mean barrier diverge. One finds $\lim\limits_{t\rightarrow t_{K}}\overline{%
F^{\ddagger }}\propto (t-t_{K})^{-2}$ and $\lim\limits_{t\rightarrow
t_{K}}R\propto (t-t_{K})^{-1}$. Since the droplet interface $\xi $ remains
finite as $t\rightarrow t_{K}$, the thin wall approximation is well
justified close to the Kauzmann temperature. On the other hand, $R$ and $\xi 
$ become comparable for temperatures close to $T_{A}$ and the thin wall
approximation breaks down. We see that replica Landau functional calculation predicts
a rather diffuse droplet near the laboratory $T_g$. Combining $R\propto (t-t_{K})^{-1}$ and $%
\overline{s_{c}}\propto (t-t_{K})$, we obtain $\nu =1$ for the exponent that
relates the droplet size $R$ and the configurational entropy density: $%
R\propto s_{c}^{-\nu }$.

In Ref. \onlinecite{Moore06} it was demonstrated that for a similar model the interface free
energy is exponentially small for a large system and in any finite
dimension the one step replica symmetry
breaking state does not exist.  As was already pointed out in Ref.  \onlinecite{Moore06} , our
approach is very different in scope and in its conclusions. While Ref.  \onlinecite{Moore06} is
concerned with the absence of replica symmetry breaking in the ultimate
equilibrium state, our conclusions are relevant for the nonequilibrium
sitation. Since we obtain a finite barrier height of the frozen mean field
solution, our result offers a mechanism for equilibration on time scales $%
t\gg \overline{\tau }$ demonstrating that our approach and the conclusions
of Ref.  \onlinecite{Moore06} are consistent. 

\subsection{barrier fluctuations}

To analyze the barrier fluctuations we start from the barrier for a given
density configuration $\rho $ : 
\begin{equation}
F_{p_{c},\rho }^{\ddagger }=F_{p_{c},\rho }-F_{q^{\ast },\rho }
\label{Fbarr}
\end{equation}%
where $F_{p_{c},\rho }=-T\log Z_{p_{c},\rho }$ is determined by the
constrained partition function of Eq.\ref{Zprho} and $F_{q^{\ast },\rho }$
the corresponding energy with homogeneous overlap, i.e. \ $F_{q^{\ast },\rho
}=-T\log Z_{p_{c}(r\rightarrow \infty )\rightarrow q^{\ast },\rho }$. When
we analyze variations of \ activation barriers we need to keep in mind that
both, ground state and transition state contributions to $F_{p_{c},\rho
}^{\ddagger }$ are statistically fluctuating and are correlated in general.
Fluctuations in the first term of Eq.\ref{Fbarr} correspond to variations
of the free energy of the localized instanton. On the other hand,
fluctuations in the second term correspond to variations of the homogeneous
background. In principle cancellations between both occur that are properly
included in the analysis of $F_{p_{c},\rho }^{\ddagger }$ defined above.

Barrier fluctuations are then characterized by the second moment 
\begin{equation}
\overline{\delta F^{\ddagger 2}}=\overline{F_{p_{c},\rho }^{\ddagger 2}}-%
\overline{F_{p_{c},\rho }^{\ddagger }}^{2}  \label{Bfl}
\end{equation}%
where the average is, just as for the analysis of the mean barrier, with
respect to the density configuration $\rho $. The second term\ in Eq.\ref%
{Bfl} is the square of the mean activation barrier $\overline{F^{\ddagger }}$
and was determined in the previous section. Thus, we can concentrate on
first term. \ Using Eq.\ref{Fbarr} it follows that the first term in Eq.\ref%
{Bfl} consists of three terms:%
\begin{equation}
\overline{F_{p_{c},\rho }^{\ddagger 2}}=\overline{F_{p_{c},\rho }^{2}}+%
\overline{F_{q^{\ast },\rho }^{2}}-2\overline{F_{p_{c},\rho }F_{q^{\ast
},\rho }}  \label{FF}
\end{equation}%
In what follows we consider these three contributions separately. \ The
first two terms correspond to independent fluctuations of the droplet and
the homogeneous background, while the last term measures their mutual
correlations. We find that this last term in Eq.\ref{FF} is proportional to
the surface area of the instanton. Correlations between droplet and homogeneous
background terms result from their mutual interface.

The detailed analysis of the three contributions to $\overline{F_{p_{c},\rho
}^{\ddagger 2}}$ of Eq.\ref{FF} is summarized in appendix B. In what follows
we summarize the results of this derivation. For the first two terms in Eq.%
\ref{FF} it follows that

\begin{eqnarray}
\overline{F_{q^{\ast },\rho }^{2}} &=&\overline{F_{q^{\ast },\rho }}%
^{2}-2T\int d^{3}{\mathbf{r}}h[q^{\ast }]  \notag \\
\overline{F_{p_{c},\rho }^{2}} &=&\overline{F_{p_{c},\rho }}^{2}-2T\int d^{3}%
\mathbf{r}h[p_{c}\left( \mathbf{r}\right) ].  \label{fttb}
\end{eqnarray}%
The first expression gives fluctuations of the configurational entropy for
the homogeneous problem, while the second one describes the energy
fluctuations of the configurations with spatially heterogeneous overlap.

The result for the homogeneous problem, $\overline{F_{q^{\ast },\rho }^{2}}$
can alternatively be obtained from our analysis of the configurational heat
capacity given in Eq.\ref{ffluctC}. Using the replica formulation of Eq.\ref%
{repsecd} together with the explicit result of Eq.\ref{F(m)Mon} yields 
\begin{equation}
\overline{F^{2}}-\overline{F}^{2}=-2VT\left( \frac{t}{2}q^{\ast 2}-\frac{w+u%
}{3}q^{\ast 3}+\frac{y}{4}q^{\ast 4}\right)  \label{flMon}
\end{equation}%
where $V$ is a volume. If we now recall the definition of $h[q]$ Eq.\ref%
{ham2}, one readily observes that the first equation in Eq.\ref{fttb}
coincides with Eq. (\ref{flMon}).

The derivation of the third contribution $\overline{F_{p_{c},\rho
}F_{q^{\ast },\rho }}$ to $\overline{F_{p_{c},\rho }^{\ddagger 2}}$ in Eq.%
\ref{FF}, which describes the energy correlations between the instanton and
environment, is also performed in the Appendix B. The result is: 
\begin{eqnarray}
\overline{F_{p_{c},\rho }F_{q^{\ast },\rho }} &=&\overline{F_{p_{c},\rho }}%
\times \overline{F_{q^{\ast },\rho }}-T\int d^{3}\mathbf{r}\left(
2h[p_{c}\left( \mathbf{r}\right) ]\right.   \notag \\
&&\left. +up_{c}^{3}(\mathbf{r})-uq^{\ast }p_{c}({\mathbf{r}})\right) .
\label{flcorr}
\end{eqnarray}%
Combining Eqs.\ref{fttb} with Eq.\ref{flcorr}, we finally obtain for the
second moment of the barrier fluctuations: 
\begin{equation}
\frac{\overline{\delta F^{\ddagger 2}}}{2T}=\int d^{3}r\left(
h[p_{c}]-h[q^{\ast }]+up_{c}^{3}({\mathbf{r}})-uq^{\ast }p_{c}^{2}({\mathbf{r%
}})\right) .  \label{barfl}
\end{equation}%
This result for the second moment of the barrier fluctuations 
depends on the value of the homogeneous overlap $q^{\ast }$ of Eq.\ref{qstar}
as well as the instanton solution $p_{c}({\mathbf{x}})$. The thin wall limit
result for $p_{c}({\mathbf{x}})$ is given in Eq.\ref{pcthinwall}. \
Inserting these expressions into the Eq.\ref{barfl} yields 
\begin{equation}
\overline{\delta F^{\ddagger 2}}=A\left( R^{3}+\rho R^{2}\right) ,
\label{Fddag}
\end{equation}%
where $R$ is the radius of the droplet. In the thin-wall approximation, 
the explicit expressions for the coefficient $A$ is 
\begin{widetext}
\begin{equation}\label{At}
\frac{A}{k_{B}TE_{0}}=\frac{2\pi y}{3a_0^3}\left[
-\left( 1+\frac{u}{w}\right) 
(q^{\ast 3}-q_{1}^{3})+q^{\ast 4}-q_{1}^{4}+\frac{2u}{y}%
(q_{1}^{3}+q^{\ast 3}-2q_{1}^{2}q^{\ast })\right] ,
\end{equation}
\end{widetext}
where the value of $q_{1}$ is computed using an effective potential $%
V(q_{1})=V(q^{\ast })$, i.e. it is the turning point of the intstanton, see
Fig.3. Deriving (\ref{At}) we also made the following choice for the parameters
$q^\ast(t_K)=2w/3y=1$ (see Appendix A for more details). This choice is of particular
convenience as it allows one to express the resulting expressions in terms of the 
ratio $t/t_K$ and $u$. 
The length scale $\rho (t)$ can be compactly as follows:%
\begin{equation}\label{rho}
\begin{split}
\rho (t)&=\frac{4\pi t_K\xi(t)}{a_{0}^{3}}\left( \frac{k_{B}TE_{0}}{A}\right)
\times\\&\times\left[\frac{2uq^{\ast}}{t_K}c_2(t)-\left(1+\frac{4u}{3t_K}\right)c_3(t)+c_4(t)\right],
\end{split}
\end{equation}%
where $c_n(t)$ are functions of the ratio $t/t_K$ only and are defined via 
\begin{equation}\label{cn}
c_n(t)=\int\limits_0^\infty \left[q^{\ast n}-p_c^n(z)\right]dz
\end{equation}
We have computed these integrals numerically for $n=2,3$ and $4$.  
Temperature dependence of the surface length scale $\rho(t)$ is shown on Fig. \ref{fig5}. 

Before we proceed with the analysis of our results, it is instructive to write down the 
expressions for the coefficient $A$ and $\rho(t)$ for temperatures
close to $t_K$. When $t\simeq t_K$, one readily finds $q_1\simeq t_K(t-t_K)/y$, 
so that expanding expression in the square brackets (\ref{At}) up to the 
linear order in $t-t_K$ we have 
$A(t)\simeq (k_BTE_0/a_0^3)[A_0+A_1\cdot(t-t_K)]+{\cal O}((t-t_K)^2)$
with
\begin{equation}
\begin{split}
&A_0=\frac{8\pi}{9}u, \quad 
A_1=4\pi\left(\frac{2u}{t_K}+1\right),
\end{split}
\end{equation}
where once again we employed $q^*(t_K)=1$ (see Appendix A). Thus we find that close to 
$t_K$ the bulk contribution to the barrier fluctuations is governed solely by the value of the 
phenomenological parameter $u$. Expression for 
$\rho(t\simeq t_K)=\rho_K+\rho_{1K}\cdot(t-t_K)$
can be derived straightforwardly using our results above together with the expansion for the functions $c_n(t)$ and $\xi(t)\simeq 2/\sqrt{t_K}+5(t-t_K)/t_K^{3/2}$. Here we provide the expression for the value of $\rho(t=t_K)$:
\begin{equation}
\rho(t_K)=\frac{9t_K^{1/2}}{u}\left[c_{4K}-c_{3K}+\frac{2u}{t_K}\left(c_{2K}-\frac{2}{3}c_{3K}\right)\right],
\end{equation}
where we have used the notations $c_{nK}\equiv c_n(t_K)$. We have found that the 
coefficients $c_{nK}$ have the following values:
$c_{2K}\simeq 1.25$, $c_{3K}\simeq 1.18$ and 
$c_{4K}\simeq 1.11$. 
%%%%%%%%%% Fig6: Barrier fluctuations
\begin{figure}[h]
\includegraphics[width=2.8in]{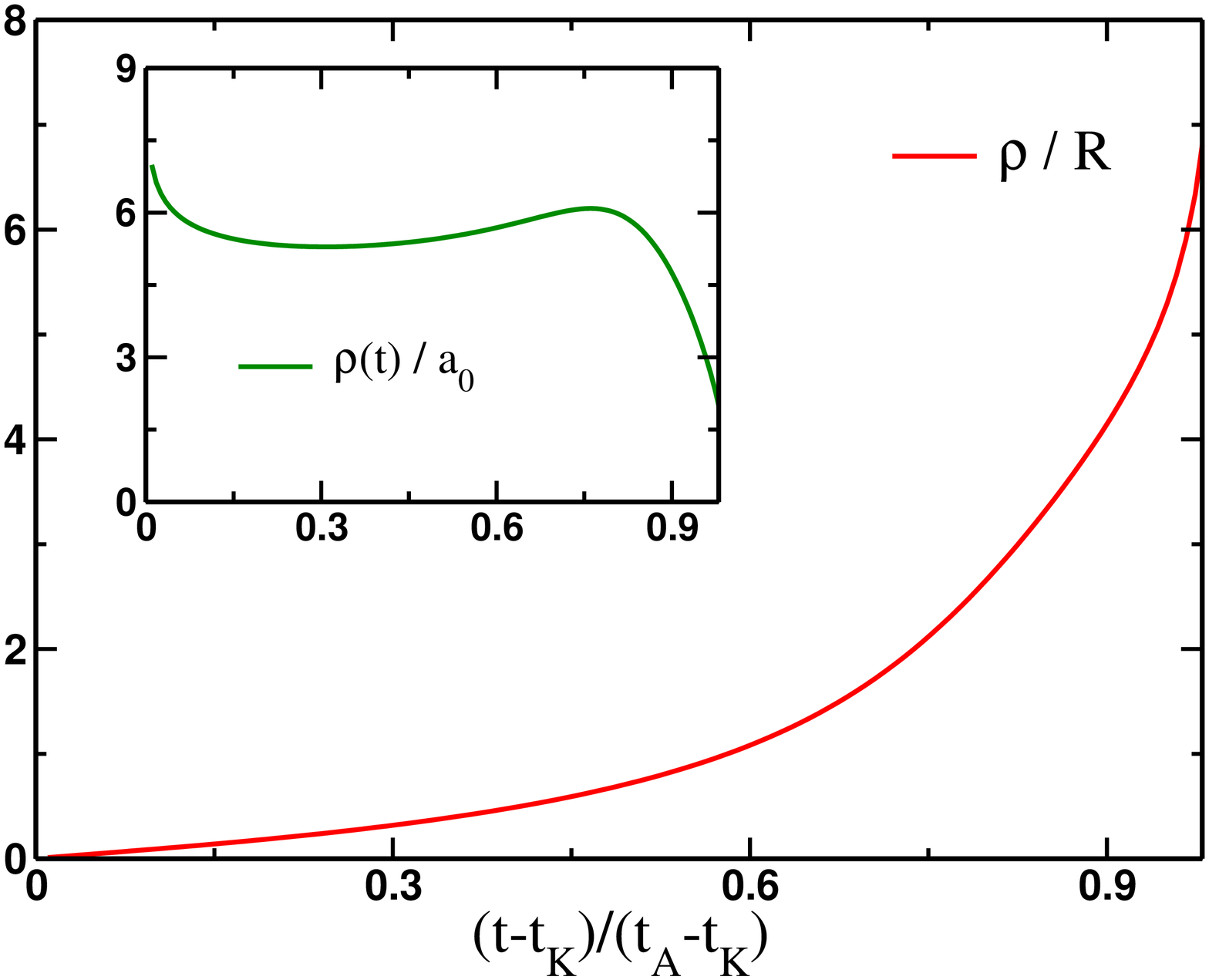}
%\includegraphics[width=3.0in]{Fig5_rho.pdf} 
%\centerline{\psfig{file=Fig5_rho.eps,height=6cm,width=8cm,angle=-90}}
\caption{Temperature dependence of the length scale $\protect\rho (t)$ 
describing the contribution to the barrier fluctuations
due to surface energy fluctuations of the structural droplets. We also show the temperature
dependence of the ratio $\rho/R$ ($R$ is the size of the droplet).}
\label{fig5}
\end{figure}
%%%%%%%%%% End of Fig6 %%%%%%%%%%%%%%%%%%

%%%%%%%%%% Fig7: Barrier fluctuations
\begin{figure}[h]
\includegraphics[width=2.8in]{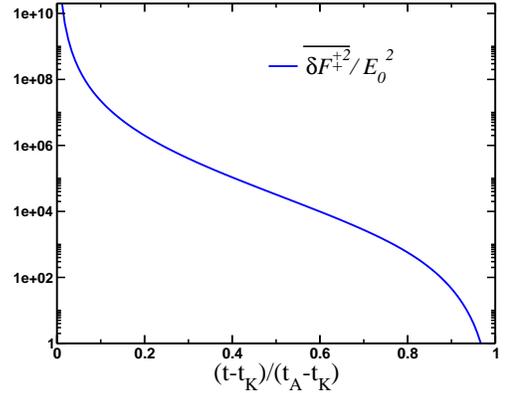}
%\includegraphics[width=3.0in]{Fig6_barfl.pdf} 
%\centerline{\psfig{file=Fig6_barfl.eps,height=6cm,width=8cm,angle=-90}}
\caption{Temperature dependence of the energy barrier fluctuations.
scale $\protect\rho (t)$ (bottom
inset). The latter describes the contribution to the barrier fluctuations
due to surface energy fluctuations of the structural droplets.}
\label{fig6}
\end{figure}
%%%%%%%%%% End of Fig7 %%%%%%%%%%%%%%%%%%

%%%%%%%%%% Fig8: exponents 
\begin{figure}[h]
\includegraphics[width=2.8in]{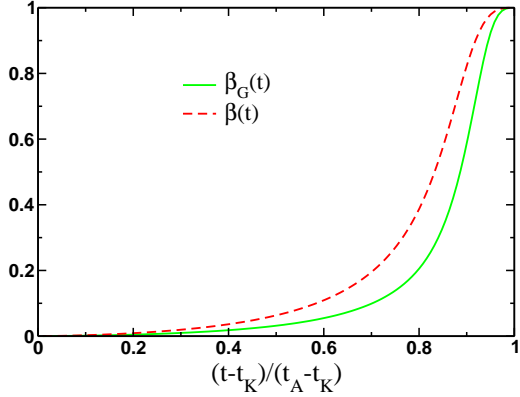}
%\includegraphics[width=3.0in]{Fig7_betas.pdf} 
%\centerline{\psfig{file=Fig7_betas.eps,height=6cm,width=8cm,angle=-90}}
\caption{Temperature dependence of the exponent 
$\protect\beta_G$ governed by the fluctuations of the isolated droplets 
together with exponent $\beta$ corrected by the effects of the droplet-droplet interactions.}
\label{fig7}
\end{figure}
%%%%%%%%%% End of Fig8
Thus, in addition to the bulk term, which behaves in a similar way as the
formulation in Ref.\cite{XW01} (see Eq.\ref{XW}), the replica theory yields
a surface term, resulting from fluctuations in the interface. In the 
inset of Fig.\ref{fig5} we show the $T$-dependence of $\ $the length scale $\rho (t)$,
 demonstrating that the surface term becomes gradually more important as
the temperature increases. The value of $\rho(t_K)$ remains non-singlar, so that 
the volume term is dominant. This is consistent with the view that the droplet
interface close to $T_{K}$ is smaller then the droplet radius, while
both are comparable as $T$ approaches $T_{A}$, \ similar to the behavior
close to a spinodal\cite{Unger84}. The resulting temperature dependence of 
$\rho(t)$ is shown on Fig. \ref{fig5}, 
$\overline{\delta F^{\ddagger 2}}$ is shown in Fig. \ref{fig6} and the exponent of the streched exponential relaxation $\beta_G$ shown in Fig. \ref{fig7}.

The plots are constructed by setting the parameters of the
theory to re-produce experimentally relevant values of $T_{K}$, $T_{g}$ and $%
T_{A}$ as well as configurational entropy above the glass transition using 
$o-$terphenyl as the example of the glass transition 
(see Appendix A for details). There is one free parameter left to be fixed,
either $y$ or $w=3y/2$. We have determined the value of $y$ in the comparison 
with the experimental value of the mean barrier at $T=T_{g}$ given by $\overline{%
F^{\ddagger }}=T_{g}s_{c}(T_{g})$ with the theoretically derived expression (%
\ref{Fdagger}). This procedure yields $y=1.82$, $w=2.73$ and $u=0.385$. As we can see
from Fig. \ref{fig7}, $\beta_G(T_{g})\simeq 0.12$ , a value
significantly reduced compared to the exponential behavior where $\beta =1$.
This value for $\beta_G$ is lower than the experimentally found values of $%
\beta _{exp}(T=T_{g})\simeq 0.5$. This discrepancy may signal that the present
approach overestimates the strength of the energy barrier fluctuations an
effect that may be related with non-Gaussian fluctuations of the activation
barriers. It is however clear that these may be described 
as instanton interaction effects as discussed by Xia and Wolynes\cite{XW00}. 
%%%%%%%%%%%%%%%%%%%%%%
Our analysis yields a Gaussian distribution of barriers%
\begin{equation}
P_{G}\left( F^{\ddagger }\right) =\frac{1}{\sqrt{2\pi \overline{\delta
F^{\ddagger 2}}}}\exp \left( -\frac{\left( F^{\ddagger }-\overline{%
F^{\ddagger }}\right) ^{2}}{2\overline{\delta F^{\ddagger 2}}}\right) 
\end{equation}%
which yields the following estimatefor the exponent of the streched
exponential relaxation  
\begin{equation}
\beta _{G}=\left( 1+\overline{\delta F^{\ddagger 2}}/\left( k_{B}T\right)
^{2}\right) ^{-1/2}.
\end{equation}%
Our results for the temperature dependence of $\beta _{G}$ are shown in
Fig.\ref{fig7}. The calculation that led to this result was based on the assumption
of single instanton events, i.e. \ entropic droplets were assumed to be
diluted. Obviously this is not addressing the fact that distict droplets
interact. The mosaic picture underlying the RFOT theory is clearly based
upon the view that such droplet-droplet interactions occur and are in fact
crucial. The impact of droplet-droplet interactions for the distribution of
barriers was analyzed in Ref.\cite{XW00}.   Here it was pointed out that, as soon
as the droplet size becomes larger than the size $R$ of a mean droplet
(Eq.55 in our theory), boundary effects will limit the size of a droplet,
leading to a cut off of the distribution function. This leads to the
modified distribution function 
\begin{equation}
P\left( F^{\ddagger }\right) =\left\{ 
\begin{array}{cc}
\frac{1}{2}P_{G}\left( F^{\ddagger }\right)  & F^{\ddagger }\leq \overline{%
F^{\ddagger }} \\ 
\frac{1}{2}\delta \left( F^{\ddagger }-\overline{F^{\ddagger }}\right)  & 
F^{\ddagger }>\overline{F^{\ddagger }}%
\end{array}%
\right. 
\end{equation}%
which leads to a reduction of the mean square deviation of the barriers by a
factor $1/4$, compared to the Gaussian distribution. This corrects the
exponent of the streched exponential relaxation to%
\begin{equation}
\beta =\left( 1+\frac{1}{4}\overline{\delta F^{\ddagger 2}}/\left(
k_{B}T\right) ^{2}\right) ^{-1/2}
\end{equation}%
where $\overline{\delta F^{\ddagger 2}}$ is still the result, Eq.63, of the
dilute droplet calculation. In Fig.\ref{fig7} we compare $\beta _{G}$ and $\beta $.
Obviously, the reduction of the barrier fluctuation width due to
droplet-droplet coupling leads to an increase of the exponent $\beta $. 
We mention however 
the resulting value of $\beta$ at the glass transition is $\beta(T_g)\simeq 0.23$. 
This suggests the present Landau functional indeed over-estimates the fluctuation effects. 
This probably implies that the diffuseness 
of the droplet from the Landau functional is also overestimated. 
The fluctuations do vanish at $T_A$, consistent with the emergence of simple exponential
dynamics at a significantly high temperature. We did not attempt to vary the parameters of our model to reach perfect agreement with experiments for OTP, as the simplified Landau functional is clearly an oversimplification of the physics of real glasses. It does however demonstrate the generic and qualitative trends for barrier fluctuations in the RFOT theory.
%%%%%%%%%%%%%%%%%%%%%% end of Joerg's insert $$$$$$$$$$$$$$$

\begin{figure}[h]
\includegraphics[width=2.8in,angle=-90]{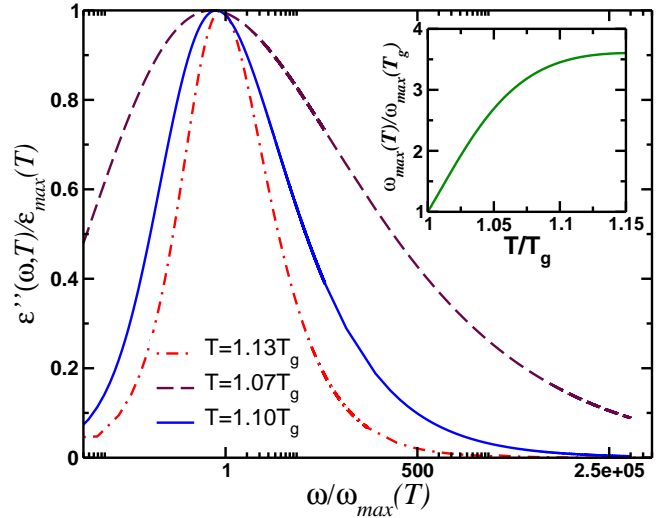}
%\includegraphics[width=3.0in]{Fig8_diel.pdf} 
%\centerline{\psfig{file=Fig8_diel.ps,height=6cm,width=8cm,angle=-90}}
\caption{Frequency dependence of the imaginary part of the dielectric
susceptibility above the glass transition. The broadening of the peak at the
glass transition signals an appearance of wide spectrum of excitations.}
\label{fig8}
\end{figure}

Finally, in Fig. \ref{fig8} \ we show \ the
dielectric finction $\varepsilon ^{\prime \prime }\left( \omega ,T\right) $
as finction of frequency for different temperatures. The peak in $%
\varepsilon ^{\prime \prime }(\omega ,T)$ dramatically broadens as
temperature approaches the glass transition temperature, $T_{g}$, 
as predicted from the static barrier distributions. 
%%%%%%%%%% Fig.9: dielectric constant
%%%%%%%%%% End of Fig.9
For this figure we used the second moment $\overline{\delta F^{\ddagger 2}}$
of the barrier distribution function and then assumed that $p\left(
F^{\ddagger }\right) $ is Gaussian. In case of a Gaussian distribution, all
moments can be expressed in terms of the second moment 
\begin{equation}
\overline{\delta F^{\ddagger k}}=\mu _{k}^{G}\left( \overline{\delta
F^{\ddagger 2}}\right) ^{k/2}
\end{equation}%
where $\mu _{k}^{G}=0$ for $k$ odd and $\mu _{k}^{G}=2^{k}\Gamma \left( 
\frac{k+1}{2}\right) /\sqrt{\pi }$ for $k$ even. In particular it holds for
higher order moments $\mu _{3}^{G}=0$, $\mu _{4}^{G}=3$, $\mu _{5}^{G}=0$,
and $\mu _{6}=15$ etc.

We can explicitly analyze higher moments of this distribution and thereby demonstrate
explicitly that $p\left( F^{\ddagger }\right) $ is Gaussian. To be precise,
we have only shown this for the first six moments but strongly suspect this to be
true in general. The calculation of higher moments are quite cumbersome but 
they formally are a straightforward generalization of the method we used to determine the
second moment $\overline{\delta F^{\ddagger 2}}$. In case of the $k$-th
moment the block structure of the matrix $Q$ of Eq.\ref{eq8} \ consists of $%
\left( k+1\right) \times \left( k+1\right) $ blocks where the total
dimension of $Q$ is $\left( n+\sum_{i=1}^{k}m_{k}\right) \times \left(
n+\sum_{i=1}^{k}m_{k}\right) $. The rather tedious matrix algebra is then
most easily analyzed using computer algebra software, such as Mathematica. Thus,
we only list the results here. We analyzed all moments%
\begin{equation}
\overline{\delta F^{\ddagger k}}=\int \left( F^{\ddagger }-\overline{{F}%
^{\ddagger }}\right) ^{k}p\left( F^{\ddagger }\right) dF^{\ddagger }
\end{equation}%
up to $k=6$ of . We find 
\begin{equation}
\overline{\delta F^{\ddagger 3}}=0.
\end{equation}%
In case of the fourth moment we analyze:%
\begin{eqnarray}
\overline{\delta F^{\ddagger 4}} &=&\overline{(F^{\ddagger }[\rho ]-%
\overline{{F}^{\ddagger }})^{4}}  \notag \\
&=&\overline{F^{\ddagger 4}[\rho _{2}]}-4\overline{F^{\ddagger 3}[\rho _{2}]}%
\overline{{F}^{\ddagger }}+3\overline{{F}^{\ddagger }}+6\overline{{F}%
^{\ddagger }}\cdot \overline{{F}^{\ddagger 2}},
\end{eqnarray}%
it follows 
\begin{equation}
\overline{\delta F^{\ddagger 4}}=3[2\mathcal{V}\left[ p_{c};q_{0}\right] -%
\mathcal{H}_{inh}\left[ p_{c}\right] -\mathcal{H}_{hom}(q^{\ast })]^{2}
\end{equation}%
where $\mathcal{H}_{hom}$ and $\mathcal{H}_{inh}$ are defined below Eq. \ref%
{HHH}. Evaluating this expression yields 
\begin{equation}
\overline{\delta F^{\ddagger 4}}=3\left( \overline{\delta F^{\ddagger 2}}%
\right) ^{2}
\end{equation}%
We further find that the fifth moment vanishes as well%
\begin{equation}
\overline{\delta F^{\ddagger 5}}=0
\end{equation}%
and that the sixth moment is given as 
\begin{equation}
\overline{\delta F^{\ddagger 6}}=15\left( \overline{\delta F^{\ddagger 2}}%
\right) ^{3}.
\end{equation}%
Thus, up to the sixth moment the barrier distribution of droplets is
Gaussian. Thus, dilute entropic droplets should indeed have a Gaussian
distribution of barriers, strongly suggesting that the observed non-Gaussian 
behavior of the effective barrier distribution results from droplet-droplet interactions. 

\section{Summary}

In summary, we have shown how the energy barrier fluctuations $\overline{%
\delta F^{\ddagger 2}}$ as well as higher moments of the static free energy barrier
distribution function $p\left( F^{\ddagger }\right) $ can be computed using the
replica Landau theory of Ref.\cite{Franz95} and use the replica instanton
theory of Refs.\cite{Franz05-1,Dzero05} . We have generalized the replica
formalism to determine barrier fluctuations in addition to the determination
of the mean barrier $\overline{F^{\ddagger }}$. \ Barrier fluctuations 
$\overline{\delta F^{\ddagger 2}}$ consist of a dominating bulk contribution at 
temperatures close to the Kauzmann temperature
along with a contribution stemming from fluctuations in the interface. The latter dominate
the barrier fluctuations well above the Kauzmann temperature. 
Fluctuations of the surface energy enter prominently in the recent work of 
Biroli et al. \cite{Biroli08} on directly computing point-to-set correlations in liquids. 
It is very nice to see that the effect they found naturally emerges from the replica
instanton calculation framework. It is interesting to speculate that these surface
energy fluctuations could account for the observed system specific deviations from 
the Xia-Wolynes prediction of a direct relation between the stretching exponent and
$\Delta C_v$. Chemical trends in these deviations may provide clues on this. 
$\overline{\delta F^{\ddagger 2}}$ increases as the typical droplet size
increases, i.e. as the temperature is lowered towards $T_{K}$. Clearly, our
theory only applies until the system falls out of equilibrium at the
laboratory glass temperature $T_{g}$ with $T_{g}>T_{K}$. \ In the framework of
the present instanton approach,\ the energy barrier distribution function
turns out to be Gaussian. \ The skewness of the observed relaxation time 
distribution is, in our view, an effect due to the interaction of spatially overlapping
instantons. It can be captured, at least in spirit, by an extended mode coupling approach,
incorporating instantons.

\section{Acknowledgement}

This research was supported by the Ames Laboratory, operated for the U.S.
Department of Energy by Iowa State University under Contract No.
DE-AC02-07CH11358 (M.D. and J. S.), a Fellowship of the Institute for
Complex Adaptive Matter (M.D.), and the National Science Foundation grants
CHE-0317017 (P. G. W.) and DMR 0605935 (M.D.) The authors also thank 
the Aspen Center for Physics, where part of this work was performed.

\appendix

\section{Estimate of the model parameters for o-terphenyl}

The Landau functional is quite convenient for formal use but there are variety 
of ways that it may be
mapped onto real fluids. In this appendix we show how
the model parameters of the Landau replica potential, Eqs.\ref{ham1}
\ and \ref{ham2}, can be found by fitting the results of the mean field
theory to experimental data. We use the concrete example o-terphenyl (OTP), a well
known glassforming material. Values for the effective dynamical freezing temperature
and the Kauzmann temperature of OTP are $T_{A}=285\mathrm{K}$\cite{Gottke01}
and $T_{K}=202.7\mathrm{K}$\cite{Chang72,Greet70}. On the other hand, a
typical value for the glass temperature is\cite{Greet67} $T_{g}\simeq
243...246\mathrm{K.}$ Here, the glass transition temperature is the
temperature at which the viscosity reaches a value of $10^{13}$\textrm{%
g/(cm s)}, which is also where the mean
relaxation time reaches values of $10^{2}\mathrm{s}$. The melting
temperature of OTP is $T_{m}=329\mathrm{K}$ with an entropy jump at melting
of $\Delta S_{m}=52.28\frac{\mathrm{J}}{\mathrm{molK}}$.

The configurational entropy at $T_{g}$ and $T_{A}$ are $S_{c}\left(
T_{g}\right) =21.5\frac{\mathrm{J}}{\mathrm{molK}}=2.\,\allowbreak 59\mathrm{%
R}$ and $S_{c}\left( T_{A}\right) =39\frac{\mathrm{J}}{\mathrm{molK}}%
=4.\,\allowbreak 69\mathrm{R}$, respecticely\cite{Chang72,Johari00}, where $%
\mathrm{R}$ is the molar gas constant. To obtain the entropy per spherical object or
bead one has to divide these results by $n_{B}=3.7$, the number of beads as defined
by the procedure
in Ref.\cite{Lubchenko04a}. This yields $s_{c}\left( T_{g}\right) \equiv 
\frac{S_{c}\left( T_{g}\right) }{n_{B}}=0.7R$ as well ws $s_{c}\left(
T_{A}\right) =1.\,\allowbreak 26R$.

The heat capacity change at the glass transition is\cite{Wang02} 
\begin{equation}
\Delta c_{p}\left( T_{g}\right) =111.27\frac{\mathrm{J}}{\mathrm{molK}}=0.48%
\frac{\mathrm{J}}{\mathrm{gK}}
\end{equation}%
where we used $1\mathrm{g}=\ 4.31\times 10^{-3}\mathrm{mol}$. The specific
and molar volume of the system are\cite{Greet67}: 
\begin{eqnarray}
v\left( T_{A}\right) &=&0.918\frac{\mathrm{cm}^{3}}{\mathrm{g}}=213.2\frac{%
\mathrm{cm}^{3}}{\mathrm{mol}}  \notag \\
v\left( T_{g}\right) &=&0.893\frac{\mathrm{cm}^{3}}{\mathrm{g}}=207.3\frac{%
\mathrm{cm}^{3}}{\mathrm{mol}}.
\end{eqnarray}%
The typical volume per particle $V/N=\frac{4\pi }{3}l_{0}^{3}$ is then
characterized by the length $l_{0}$. It follows $V/N=344.\,\allowbreak 23{%
\ddot{A}}^{3}$, such that $l_{0}=4.35{\ddot{A}}$. In comparison, the van der
Waals radius of OTP was given as\cite{Edwards70} $r_{\mathrm{vdW}}=3.7{\ddot{%
A}}$.

If we introduce the energy and length scales into the problem, the mean
field theory results obtained from the Hamiltonian \ Eqs.\ref{ham1} \ and %
\ref{ham2} are 
\begin{eqnarray}
q_{0}\left( T_{K}\right) &=&\frac{2w}{3y}\   \notag \\
T_{A} &=&T_{0}\left( 1+\frac{E_{0}}{T_{0}}\frac{w^{2}}{4y}\right)  \notag \\
T_{K} &=&T_{0}\left( 1+\frac{E_{0}}{T_{0}}\frac{2w^{2}}{9y}\right)  \notag \\
T_{A}s_{c}\left( T_{A}\right) &=&\frac{4\pi }{3}l_{0}^{3}\frac{E_{0}}{%
a_{0}^{d}}\frac{w^{4}}{192y^{3}}  \notag \\
s_{c}\left( T\simeq T_{K}\right) &\simeq &\frac{4\pi }{3}l_{0}^{3}\frac{%
E_{0}w}{6T_{K}a_{0}^{d}}\left( \frac{2w}{3y}\right) ^{3}\frac{t-t_{K}}{t_{K}}
\notag \\
T_{K}c_{c}\left( T_{K}\right) &=&\frac{4\pi }{3}l_{0}^{3}\frac{E_{0}}{%
a_{0}^{d}}\frac{2u}{3}\left( \frac{2w}{3y}\right) ^{3}  \label{expr}
\end{eqnarray}%
The order parameter at $T_{K}$ should be large of order unity. We chose $%
q_{0}\left( T_{K}\right) =1$ which gives $w=\frac{3}{2}y$. Using the relation $%
T_{A}-T_{K}=\frac{w^{2}}{36y}E_{0}$ this yields
\begin{equation}
E_{0}w=1975.\,\allowbreak 2\mathrm{K}=9.\,\allowbreak 74T_{K}  \label{Ew}
\end{equation}%
If we further make the reasonable choice of $a_{0}=0.87r_{vdW}~$for
the length scale $a_{0}$, $s_{c}\left( T_{A}\right) $ of Eq.\ref{expr} gives
precisely the experimental value listed above. We can also determine the
change in heat capacity $\Delta c_{p}\left( T_{g}\right) $ from the same
parameters.  Eq. \ref{Ew} yields the result that 
\begin{equation}
\frac{t-t_{K}}{t_{K}}\simeq 0.31\frac{T-T_{K}}{T_{K}}
\end{equation}%
so that we obtain for the temperature dependence of the configurational entropy: 
\begin{equation}
s_{c}\left( T\right) \simeq \Delta c_{p}\left( T_{g}\right) \frac{T-T_{K}}{%
T_{g}}  \label{dcp}
\end{equation}%
where the mean field result for the slope is $\Delta c_{p}\left(
T_{g}\right) =\ 4.36R$. This compares reasonably well with the experimenal
value $\Delta c_{p}\left( T_{g}\right) =\frac{\Delta C_{p}\left(
T_{g}\right) }{n_{B}}=\allowbreak 3.\,\allowbreak 62R$. Note, however, that $\Delta c_{p}$ as
given by Eq.\ref{dcp} is different from the configurational heat capacity $%
c_{c}$, Eq. (\ref{ConfHeatCap}). It holds $s_{c}\simeq \frac{w}{4u}c_{c}\left( t_{K}\right) \ \frac{%
t-t_{K}}{t_{K}}$, i.e. $\frac{w}{4u}c_{c}\left( t_{K}\right) 0.253=\Delta
c_{p}\left( T_{g}\right) $, i.e. both are comparable provided $u\simeq 0.1w$, a value
we will assume in what follows thus ensuring thermodynamic consistency from the two 
fluctuation formulas. In summary, the existing thermodynamic
measurements can determine several parameters and allow a consistent
description of several thermodynamic data. Still, one free parameter, either
$w$ or $y$ remains since only the ratio $w$ to $y$ is determined. We use this freedom to
obtain the correct value of the mean barrier at the glass temperature.
Finally, we note that the above result also \ imply that $\frac{t_{g}-t_{K}}{%
t_{K}}=0.06\,\allowbreak 6$ i.e. about half of $\frac{t_{A}-t_{K}}{t_{K}}%
=0.125$. It is worth mentioning that the values of the Landau parameters 
are consistent with the random first order framework as assumed. 
\begin{table}[ht]
\caption{Parameters for the Landau functional}
\centering
\begin{tabular}{cccccc}
\hline\hline
 $\frac{a_0}{r_{dvW}}$ & $\frac{t_A}{t_K}$ & $\frac{t_g}{t_K}$ & $w$ & $u$ & $y$  \\ [1ex]
\hline
 0.1 & 1.125 & 1.066 & 2.73 & 0.385 & 1.82 \\ [1ex]
 \hline
 \end{tabular}
 \label{summary}
 \end{table}
Finally, we summarize our results for the model parameters by fitting the experimentally 
relevant quantities to the data obtained for OTP in Table \ref{summary}.

\section{Details of the replica calculation for barrier fluctuations}

In this appendix we summarize the derivation of the various contributions to
the barrier fluctuation\ given in Eq.\ref{FF}. We start our calculation with
the third term in Eq.\ref{FF}. The other two terms are simpler to determine
and can be obtained as specific limits of the third one. It holds: 
\begin{eqnarray}
\overline{F_{p_{c},\rho _{2}}F_{q^{\ast },\rho _{2}}} &=&T^{2}\frac{\int 
\mathcal{D}\rho {e}^{-\beta {H}[\rho ]}\log {Z}_{p_{c},\rho }\log {Z}%
_{q^{\ast },\rho }}{\int \mathcal{D}\rho {e}^{-\beta {H}[\rho ]}}  \notag \\
&=&\lim\limits_{m_{1,2}\rightarrow {0}}\left( \mathcal{Y}_{12}-\mathcal{Y}%
_{1}-\mathcal{Y}_{2}\right) .  \label{eq5}
\end{eqnarray}%
where we introduced 
\begin{eqnarray}
\mathcal{Y}_{12} &=&T^{2}\frac{\int \mathcal{D}\rho {e}^{-\beta {H}[\rho
]}Z_{p_{c},\rho }^{m_{1}}Z_{q^{\ast },\rho }^{m_{2}}+Z}{m_{1}m_{2}Z}\  
\notag \\
\mathcal{Y}_{1} &=&\ \frac{T^{2}\int \mathcal{D}\rho {e}^{-\beta {H}[\rho
]}Z_{p_{c},\rho }^{m_{1}}}{m_{1}m_{2}Z}  \notag \\
\mathcal{Y}_{2} &=&\ \frac{T^{2}\int \mathcal{D}\rho {e}^{-\beta {H}[\rho
]}Z_{q^{\ast },\rho }^{m_{2}}}{m_{1}m_{2}\int \mathcal{D}\rho {e}^{-\beta {H}%
[\rho ]}}
\end{eqnarray}%
where $Z=\int \mathcal{D}\rho {e}^{-\beta {H}[\rho ]}$. To analyze the first
term in Eq.\ref{eq5} we use again the replica trick and write: 
\begin{eqnarray}
\mathcal{I}_{12} &=&\lim\limits_{m_{1,2}\rightarrow {0}}\mathcal{Y}_{12} 
\notag \\
&=&\lim\limits_{n,m_{1,2}\rightarrow {0}}\int \mathcal{D}\varphi _{1}\frac{{e%
}^{-\beta {H}[\varphi _{1}]}Z_{p_{c},\rho }^{m1}Z_{q^{\ast },\rho }^{m_{2}}+Z%
}{m_{1}m_{2}\left( \int \mathcal{D}\varphi {e}^{-\beta {H}[\varphi ]}\right)
^{n-1}}  \notag \\
&=&\lim\limits_{m_{1},m_{2}\rightarrow {0}}\lim\limits_{n\rightarrow {0}}%
\frac{\mathcal{I}_{p_{c};q^{\ast }}^{(n;m_{1},m_{2})}}{m_{1}m_{2}}
\label{eq6}
\end{eqnarray}%
Here we introduced a quantity $\mathcal{I}_{p_{c};q^{\ast
}}^{(n;m_{1},m_{2})}$ that can be determined from an $n+m_{1}+m_{2}$
replicated problem with order parameter ${\mathbf{Q}}$: 
\begin{equation}
{\mathbf{Q}}=\left( 
\begin{array}{ccc}
r & p & q \\ 
p^{T} & u & \psi \\ 
q^{T} & \psi ^{T} & v%
\end{array}%
\right) .  \label{eq8}
\end{equation}%
It holds 
\begin{eqnarray}
\mathcal{I}_{p_{c};q_{0}}^{(n;m_{1},m_{2})} &=&\int \mathcal{D}{\mathbf{Q}}%
e^{-\beta {H}[{\mathbf{Q}}]}\prod\limits_{{\mathbf{x},}a}\delta \left( p_{c}(%
\mathbf{x})-p_{1a}({\mathbf{x}})\right)  \notag \\
&&\times \prod\limits_{{\mathbf{x},}\alpha }\delta \left( q_{0}-q_{1\alpha }(%
{\mathbf{x}})\right) ,
\end{eqnarray}%
$H[{\mathbf{Q}}]$ has the same form as Eqs.\ref{ham1} and \ref{ham2} if one
uses the matrix ${\mathbf{Q}}$ of Eq.\ref{eq8} instead of the original
replica variable $q$. The sub-matrices of ${\mathbf{Q}}$ are given as
follows: $r$ is an $n\times {n}$ matrix, $p$ is an $n\times {m_{1}}$ matrix, 
$q_{0}$ is an $n\times {m_{2}}$ matrix, $u$ is an $m_{1}\times {m_{1}}$
matrix, $v$ is an $m_{2}\times {m_{2}}$ matrix, and $\psi $ is an $%
m_{1}\times {m_{2}}$ matrix. The matrices $p$ and $q_{0}$ obey the
additional constraints: 
\begin{eqnarray}
p_{1b}(\mathbf{x}) &=&p_{c}(\mathbf{x})~\text{for}~\forall ~b=1,...,m_{1};~~
\label{eq9} \\
q_{1\alpha } &=&q_{0}~\text{for}~\forall ~\alpha =1,...,m_{2}.  \notag
\end{eqnarray}%
which is enforced through the $\delta $-function above. We assume for a
replica symmetric instanton solution that and the elements of $\psi _{a\beta
}$ are all the same: 
\begin{equation}
\psi _{a\beta }(\mathbf{x})=\psi (\mathbf{x}).  \label{eq10}
\end{equation}%
Similary we assume $v_{\alpha \beta }=(1-\delta _{\alpha \beta })v$.

The analysis of $H[{\mathbf{Q}}]$ tedious but straightforward and yields
that it can be written as a sum of three terms:%
\begin{eqnarray}
H[{\mathbf{Q}}] &=&m_{1}\mathcal{H}_{inh}[p_{c};u]+m_{2}\mathcal{H}%
_{hom}[q_{0};v]  \notag \\
&&+m_{1}m_{2}\mathcal{V}[\psi ].  \label{HHH}
\end{eqnarray}%
Here 
\begin{eqnarray*}
\mathcal{H}_{hom}[q^{\ast };v] &=&\int d^{3}\mathbf{x}\left\{ 2h\left[
q^{\ast }\right] +\left( m_{2}-1\right) h\left[ v\right] \right. \\
&&\left. -\left( m_{2}-1\right) vq^{\ast 2}-\frac{1}{3}\left( m_{2}-1\right)
v^{3}\right\}
\end{eqnarray*}%
refers to the homogeneous problem. In addition,

\begin{eqnarray*}
\mathcal{H}_{inh}[p_{c};u] &=&\int {d^{3}\mathbf{x}}\left\{ 2h[p_{c}]\
+(m_{1}-1)h[u]\right. \\
&&-(m_{1}-1)u(\mathbf{x})p_{c}^{2}(\mathbf{x}) \\
&&\ \left. -\frac{1}{3}(m_{1}-1)(m_{1}-2)u^{3}(\mathbf{x})\right\}
\end{eqnarray*}%
is the contribution of the inhomogeneous instanton. Finally, the coupling
term between the two is given as 
\begin{eqnarray*}
\mathcal{V}[\psi ] &=&\int {d^{3}\mathbf{x}}\{2h[\psi ]+(m_{1}-1)u_{0}(%
\mathbf{x}) \\
&&+(m_{2}-1)v]\psi ^{2}(\mathbf{x})-2p_{c}(\mathbf{x})q^{\ast }\psi (\mathbf{%
x})\}.
\end{eqnarray*}%
In order to determine the various matrix elements of ${\mathbf{Q}}$ defined
in Eq.\ref{eq8} we perform saddle point approximations with respect to the
numerous variables. We find $u(\mathbf{x})=r(\mathbf{x})$ which has been
calculated earlier. Similarly it follows $v=q^{\ast }$. The saddle point
equation for $\psi $ is then 
\begin{eqnarray}
p_{c}(\mathbf{x})q_{0} &=&-\nabla ^{2}\psi (\mathbf{x})+t\psi (\mathbf{x}%
)-\alpha \psi ^{2}(\mathbf{x})+y\psi ^{3}(\mathbf{x})  \label{eq24} \\
&&+[r_{0}(\mathbf{x})+q^{\ast }]\psi (\mathbf{x})  \notag
\end{eqnarray}%
Iand it follows that the solution of this equation is $\psi (\mathbf{x}%
)=p_{c}(\mathbf{x})$. With these results we are in the position to calculate
the barrier fluctuations.\bigskip

First we re-write expression Eq.(B1) by employing the saddle-point solution
as follows: 
\begin{eqnarray}
&&\overline{F_{p_{c},\rho _{2}}F_{q^{\ast },\rho _{2}}}=\lim\limits_{m_{1,2}%
\rightarrow 0}\frac{1}{m_{1}m_{2}}\left( 1-e^{-m_{2}\beta \mathcal{H}%
_{hom}[q^{\ast },q^{\ast }]}\right.  \notag \\
&&\left. -e^{-m_{1}\beta \mathcal{H}_{inh}[p_{c};r_{0},r_{1}]}+\right. 
\notag \\
&&\left. e^{-m_{1}\beta \mathcal{H}_{inh}[p_{c};r_{0},r_{1}]-m_{2}\beta 
\mathcal{H}_{hom}[q^{\ast },q^{\ast }]-m_{1}m_{2}\beta \mathcal{V}%
[p_{c}]}\right) .  \label{cross}
\end{eqnarray}%
Expanding the exponents up to the second order in $m_{i}$'s and taking $%
r_{0,1}=p_{c}$ yields the expression (57) in the text. Equations (55) can be
obtained using the same steps which lead us to (\ref{cross}). For example,
we have: 
\begin{eqnarray}
&&\overline{F_{p_{c},\rho _{2}}^{2}}=\lim\limits_{m_{1,2}\rightarrow 0}\frac{%
1}{m_{1}m_{2}}\left( 1-e^{-m_{2}\beta \mathcal{H}_{inh}[p_{c};r_{0},r_{1}]}%
\right.  \notag \\
&&\left. -e^{-m_{1}\beta \mathcal{H}_{inh}[p_{c};r_{0},r_{1}]}\right.  \notag
\\
&&\left. e^{-(m_{1}+m_{2})\beta \mathcal{H}%
_{inh}[p_{c};r_{0},r_{1}]-m_{1}m_{2}\beta \mathcal{V}_{inh}[p_{c}]}\right) ,
\end{eqnarray}%
where we introduced $\mathcal{V}_{inh}[p_{c}]=\int d^{3}{\mathbf{x}}h[p_{c}]$%
. Upon taking the limit $m_{1,2}\rightarrow 0$ and using the replica
symmetry $r_{0}=r_{1}=p_{c}$ we recover the second expression in Eq.\ref{FF}

\textbf{\ }

\section{Expression for the integration constant $z_0$}
In this Section we provide an expression for the integration constant $z_0$
which enters into Eq. (\ref{pcthinwall}) for the interface profile. First, we define the following
functions:
\begin{equation}\label{auxfun}
\begin{split}
&q_0(t)=\frac{w}{2y}\left(1+\sqrt{1-\frac{8t}{9t_K}}\right), \quad \tilde{t}=\frac{3}{2}t_K-t, \\ 
&q_m(t)=\sqrt{\frac{y}{\tilde{t}}}\left[q_0(t)-\frac{w}{3y}\right],  ~\varphi(t)=\sqrt{\frac{3q_m^2(t)-1}{2}}
\end{split}
\end{equation}
In terms of these functions the expression for $z_0(t)$ reads:
\begin{equation}
z_0(t)=\frac{1}{4}\log\left[\frac{q_m(t)+\varphi(t)}{q_m(t)-\varphi(t)}\right].
\end{equation}
Finally, we remark that the integral, which determines the length scale $\rho(t)$ (\ref{rho})
can be computed exactly using relations above. The resulting expression is quite cumbersome
and will not be listed here.

\end{document}